# Solar Neutrinos: Radiative Corrections in Neutrino-Electron Scattering Experiments

John N. Bahcall

*Institute for Advanced Study, Olden Lane, Princeton, NJ 08540*

Marc Kamionkowski

*Institute for Advanced Study, Olden Lane, Princeton, NJ 08540*
*and*
*Physics Department, Columbia University, New York, New York 10027*

Alberto Sirlin

*Department of Physics, New York University, New York, New York 10003*

**ABSTRACT**

Radiative corrections to the electron recoil-energy spectra and to total cross sections are computed for neutrino-electron scattering by solar neutrinos. Radiative corrections change monotonically the electron recoil spectrum for incident $^8$B neutrinos, with the relative probability of observing recoil electrons being reduced by about 4 % at the highest electron energies. For $p-p$ and $^7$Be neutrinos, the recoil spectra are not affected significantly. Total cross sections for solar neutrino-electron scattering are reduced by about 2 % compared to previously computed values. We also calculate the recoil spectra from $^{13}$N and $^{15}$O neutrinos including radiative corrections.

# I. INTRODUCTION

The observation of neutrino-electron scattering,

$$\nu + e \rightarrow \nu' + e', \tag{1}$$

is one of the principal techniques used to study solar neutrinos. The Kamiokande experimental team [1] has used the observed energy spectrum of recoiling electrons that results from the reaction indicated in Eq. (1) to show that: (1) the neutrinos originate from the sun since the electrons are predominantly scattered in the forward direction along the earth-sun vector; (2) the flux of $^8$B neutrinos is about one half the flux predicted by standard solar models; and (3) the energy spectrum of electron recoils is inconsistent with what is predicted [2-4] on the basis of the adiabatic MSW effect (but is consistent with the expectations for a non-adiabatic MSW effect or a large mixing angle solution of the solar neutrino problem).

Four additional solar-neutrino experiments currently under development, SNO [5], Superkamiokande [6], ICARUS [7], and BOREXINO [8], will also detect electrons scattered by solar neutrinos. The first three of these new experiments will concentrate on the $^8$B solar neutrinos and will measure more accurately (in heavy water, normal water, and argon gas, respectively) and with greatly improved statistics, the total flux of $^8$B neutrinos and the energy spectrum of the electron recoils. Because of the large counting rates expected in these new experiments, it will be possible to measure all of the quantities as a function of arrival time of the neutrinos and to look for the seasonal dependence of the neutrino fluxes, as well as to search for the day-night effects that are expected for some regions of MSW parameter space. When combined with the measurement by the SNO experiment of the flux of electron-type $^8$B neutrinos, it will be possible to infer constraints on the mixture of flavors of the neutrinos observed by the reaction in Eq. (1) since the neutrino-electron scattering cross section depends sensitively upon neutrino flavor. BOREXINO, the fourth new experiment listed above, will observe the recoil spectrum of the 0.862 MeV neutrinos produced by electron capture on $^7$Be in the solar interior, as well as the $^{13}$N and $^{15}$O solar neutrinos.

Two proposed experiments, HERON (liquid helium used as a bolometer to observe rotons [9]) and HELLAZ (high pressure helium gas used in a time-projection chamber, [10]), will make use of electron-neutrino scattering to observe the low-energy neutrinos created by the fundamental $p - p$ reaction, as well as the $^7$Be neutrino lines and the $^{13}$N and $^{15}$O neutrinos.

The thresholds in these different experiments vary from an electron kinetic energy of about 0.2 MeV (for observations of the $p - p$ neutrinos) to above 5 MeV(for observations of the $^8$B neutrinos). The maximum energy of interest for neutrinos from solar-fusion reactions is about 19 MeV [11]. In addition, neutrinos from supernovae are potentially observable with energies up to 30 MeV and beyond.

The use of electron-neutrino scattering to study solar neutrinos was first proposed by Reines and Kropp [12]; the directionality of the recoil electrons and the shape of the recoil energy spectrum was first calculated for solar neutrinos by Bahcall [13] using the then-current $V - A$ theory of beta-decay. In an early application of the Weinberg-Salam model of weak interactions, 't Hooft [14] calculated the cross sections for neutrino-electron scattering in what has since become known as standard electroweak theory. The results of 't Hooft were used by Bahcall [15] to calculate the total cross sections, the energy spectrum of the electron recoils, and the angular distribution (relative to the earth-sun vector) of the scattered electrons for all solar neutrino sources of interest.

For all of these experiments, it is natural to ask: How will radiative corrections affect the calculated cross sections that are used to interpret the observations? On dimensional grounds, the corrections are expected to be of the same order of magnitude as the fine structure constant, $\alpha \sim 1\%$. However, if the experimental precision is high the corrections may be significant. In particular, it is especially important to know if radiative corrections cause an appreciable energy-dependent modification of the electron recoil spectrum. Indeed, we recall that a measurement of the shape of the $^8$B solar neutrino energy spectrum at the earth is a practical and powerful test of standard electroweak theory, independent of solar models [16], and that different MSW solutions predict characteristic modifications of the shape of the electron recoil spectrum [2,3,11,17,18].

We evaluate the effects of radiative corrections upon the recoil energy spectrum and the total cross sections for the neutrino sources that are important for the operating and the planned experiments. We adopt the notation and results for electron-neutrino scattering that are summarized in references [11,15]. In particular, we include the one-loop electroweak and QCD corrections to the cross section for the elastic-scattering reaction in Eq. (1). We also include QED radiative corrections. We calculate the corrections to the elastic-scattering cross section and include for consistency the cross section for the inelastic process $\nu + e \to \nu' + e' + \gamma$. We assume that this final-state photon is not detected.

This paper is organized as follows. The derivation, by one of us (A.S.), of the formulae describing the radiative corrections is presented in the Appendix. The recoil energy spectra are described in Sec. II and the total scattering cross sections are given in Sec. III. We summarize and discuss the main results in Sec. IV.

## II. RECOIL ENERGY SPECTRA

In this Section, we describe the changes in the calculated recoil-energy spectra that result from including radiative corrections and an improved value, 0.2317 [19], for the $\overline{\text{MS}}$ parameter $\sin^2 \hat{\theta}_W(m_Z)$. (In references [11,15], the value $\sin^2 \hat{\theta}_W = 0.23$ was used.) The results are presented in a series of figures that give the ratio of probability distributions computed with and without the two improvements, the radiative corrections and the value of $\sin^2 \hat{\theta}_W(m_Z)$. The values for electron (muon or tau) neutrinos are displayed in dark (light) curves. The results are presented as a function of $T$, the kinetic recoil energy of the electrons.

We recall that the spectrum averaged differential cross section is given by [11,15]

$$\left\langle \frac{d\sigma}{dT} \right\rangle_T = \int_{q_{\min}}^{q_{\max}} dq \, \lambda(q) \frac{d\sigma}{dT} \, , \tag{2}$$

where $T$ is the kinetic energy of the recoil electron, $q$ the neutrino energy, $\lambda(q)$ the normalized neutrino spectrum incident at earth, and

$$q_{\min}(T) = \left\{ T + \left[ T(T + 2m_e c^2) \right]^{1/2} \right\}/2 \tag{3}$$

the minimum neutrino energy compatible with $T$.

The normalized probability, $P(T)$, for observing a recoil electron with kinetic energy between $T$ and $T + dT$ computed from Eq. (2) by including improvements [radiative corrections or precise $\sin^2 \hat{\theta}_W(m_Z)$] can be expressed in terms of the probability, $P_0(T)$, computed without improvements [11,15] as:

$$P(T) = P_0(T) \left[ 1 + \xi(T) \right]. \tag{4}$$

We present in the following subsections $\xi(T)$ versus $T$ for various solar neutrino sources: the $^7$Be lines (Sec. II A), the $p - p$ neutrinos (Sec. II B), and $^8$B neutrinos (Sec. II C).

## A. $^7$Be Electron Recoil Spectrum

Figure 1 shows in the darker lines the changes (caused by including radiative corrections and/or the more precise Weinberg angle) in the spectrum of recoil electrons produced by electron neutrinos from the 0.862 MeV $^7$Be neutrino line. The corresponding changes for $\nu_\mu - e$ scattering are shown in the lighter lines. The solid curves are the ratio—minus unity—of the probability distribution for $\nu - e$ scattering obtained using the updated value of $\sin^2 \hat{\theta}_W(m_Z)$ and including radiative corrections to the same probability distribution obtained using an older value of $\sin^2 \hat{\theta}_W(m_Z)$ with no radiative corrections. This ratio gives the net change in the probability distribution that is obtained using the formulae and constants employed in this paper compared to the values that were obtained in refs. [11] and [15]. The net changes are less than 1% over essentially all of the allowed recoil energy range. Because the radiative corrections diverge at the end-point $T_{\max}$ (see Appendix A), all spectra in this paper are displayed up to $T = 0.99\, T_{\max}$. The singularity in the expression used here is related to the infrared divergence and signals the breakdown of perturbation theory. However, the breakdown appears so close to the end point that it should be of no consequence for any realistic experiment with finite energy resolution.

Since the net changes are produced by two different effects, we also display separately in Figure 1 the effects due to the inclusion of radiative corrections and to the use of the more precise Weinberg angle. The effect of changing just $\sin^2 \hat{\theta}_W$ is shown in the dotted curves. The dotted curves are the ratios—minus unity—of the probability distributions that are obtained using $\sin^2 \hat{\theta}_W(m_Z) = 0.2317$ [19] to the probability distributions obtained using the old value of $\sin^2 \hat{\theta}_W(m_Z) = 0.23$. The dashed curves show the effects of just including radiative corrections. The dashed curves are the ratios—minus unity—of the probability distributions obtained including radiative corrections to those obtained without radiative corrections.

Each of the two effects considered here results independently in changes of less than or of order 1% in the relative probability distributions.

Fig. 2 shows results for the 0.384 MeV line that are similar to those obtained for the 0.862 MeV line.

Table I and Table II give for the $^7$Be 0.862 MeV line and the 0.384 MeV line, and for both $\nu_e - e$ and $\nu_\mu - e$ scattering, numerical values for the probability distributions that electrons recoil with kinetic energy $T$.

### B. $p - p$ Electron Recoil Spectrum

Fig. 3 shows results for neutrinos with the energy spectrum [11] characteristic of $p - p$ neutrinos. The various curves have the same meaning as in Fig. 1. For these very low-energy neutrinos, the fractional changes due to radiative corrections and to the use of the more accurate Weinberg angle are also less than and of order 1%. Table III gives, for both $\nu_e - e$ and $\nu_\mu - e$ scattering, numerical values for the probability distribution that incident $p - p$ neutrinos scatter electrons that recoil with kinetic energy $T$.

### C. $^8$B Electron Recoil Spectrum

Fig. 4 shows that the effect of radiative corrections is significant for the higher-energy neutrinos that originate from $^8$B decay. Moreover, the effect of radiative corrections is monotonic: higher energy neutrinos are suppressed relative to lower energy neutrinos. We have omitted from Fig. 4 the small, less than 0.5 %, differential effect of the change in the Weinberg angle that corresponds to the dotted curves in Fig. 1, Fig. 2, and Fig. 3.

The differential effect of radiative corrections reduces the relative probability of observing recoil electrons by as much as 4 % at the highest recoil energies. The direction of this suppression is opposite in sign to the effect that is expected from the non-adiabatic MSW solution. The highest energy neutrinos are the least suppressed in the non-adiabatic MSW solution.

Table IV gives, for both $\nu_e - e$ and $\nu_\mu - e$ scattering, numerical values for the probability distribution that incident $^8$B neutrinos scatter electrons that recoil with kinetic energy $T$.

### D. $^{13}$N and $^{15}$O Electron Recoil Spectra

In addition to the $p-p$ and $^7$Be neutrinos, there will be neutrinos produced in the CNO cycle from decays of $^{13}$N ($E_\nu \leq 1.199$ MeV) and $^{15}$O ($E_\nu \leq 1.732$ MeV). Together, these may contribute $\mathcal{O}(20\%)$ of the neutrino flux in an experiment such as BOREXINO, so it is important to include the recoil spectra from these neutrinos as well. In Table V we present for both $\nu_e - e$ and $\nu_\mu - e$ scattering, numerical values for the probability distribution that incident $^{13}$N neutrinos scatter electrons that recoil with kinetic energy $T$. In Table VI, we present the same for $^{15}$O neutrinos. We include radiative corrections and the updated value for $\sin^2 \hat{\theta}_W$ in this calculation.

In Fig. 5, we plot the recoil spectra for both electron and muon neutrinos from $p-p$, $^7$Be, $^{13}$N, and $^{15}$O neutrinos as a function of electron kinetic energy, $T$. The spectra are normalized such that the total probability is unity. The dotted curve is the recoil spectrum from $^{13}$N neutrinos and the short-dash curve is that from $^{15}$O neutrinos. The nearly flat spectrum which extends to higher energies (the solid curve) is the recoil spectrum from the 0.862-MeV $^7$Be line, while that which only extends to lower energies (the dot-dash curve) is that from the 0.384-MeV $^7$Be line. The solid curve which extends the highest at the lowest energies (the long-dash curve) is the recoil spectrum from $p - p$ neutrinos. The fractional changes in the electron recoil spectra from $^{13}$N and $^{15}$O neutrinos due to radiative corrections and the new value of $\sin^2 \hat{\theta}_W$ are no more than roughly 1%.

## III. TOTAL CROSS SECTIONS

Fig. 6 shows the total scattering cross sections as a function of neutrino energy; the notation for the different curves is the same as in Fig. 1. For $\nu_e - e^-$ scattering, the net decrease in the total cross section is about 2%, essentially independent of energy. For $\nu_\mu - e^-$ scattering, there is a net increase in the cross section of about 1.3 % relative to the numerical values given in ref. [11,15].

At first sight, it may seem puzzling that the effect of the radiative corrections on the total cross sections is essentially independent of energy, while the fractional corrections to the electron recoil spectrum are appreciably energy dependent. The explanation lies in the fact that the radiative corrections manifest themselves in two ways: (1) the QED corrections (described in Appendix B), which are energy dependent, and (2) the electroweak and QCD corrections to the coupling constants $g_L$ and $g_R$ (described in Appendix A), which are largely energy independent. The corrections to the coupling constants are generally larger and dominate the changes in the total cross sections, producing an almost constant correction to the total cross section. The effects of the corrections to the coupling constants largely cancel out of the calculation of the shape of the electron recoil spectrum, leaving an energy dependent correction to the recoil spectrum. From the appearance of Fig. 4, it would seem that the larger corrections to the recoil spectrum at larger recoil energies would have a significant impact on the total cross section at large incident neutrino energies. However, the fractional change in the recoil spectrum due to radiative corrections is large only when the recoil spectrum becomes very small (see Fig. 8.2 in ref. [11] and Table IV here).

For many purposes, it is convenient to have available numerical values for neutrino-electron scattering cross sections at specific energies. Table VII gives the computed total cross sections for neutrino energies equal to the prominent solar neutrino line energies and to typical energies in the range from 1 MeV to 60 MeV. The tabulated values were determined for $T_{min} = 0.0$ MeV.

The cross sections given in Table VII can be described approximately by the following relation:

$$\sigma(q) = \text{constant} \times \left(\frac{q}{10 \text{ MeV}}\right) 10^{-44} \text{cm}^2, \tag{5}$$

where the constant is equal to 9.2 for $\nu_e$ scattering and 1.6 for $\nu_\mu$ scattering.

The recoil spectrum for $\nu_e - e$ scattering is relatively flat. Therefore, for $\nu_e - e$ scattering, the cross sections given in Table VII can be used to estimate reasonably accurately the cross section for a specified minimum recoil electron energy. One simply multiplies the tabulated values by $(T_{max} - T_{min})/T_{max}$. This approximation is less appropriate for $\nu_\mu - e$ scattering, but still will give a useful first estimate.

In Table VIII, we list the total neutrino-electron scattering cross sections for the neutrino sources we have considered here for both electron and muon neutrinos. These are obtained by convolving the neutrino spectra from each source with the neutrino-electron scattering cross section. Radiative corrections as well as the updated value of $\sin^2 \hat{\theta}_W$ are included in the calculation.

## IV. CONCLUSIONS AND DISCUSSION

For precise solar neutrino-electron scattering experiments, radiative corrections should be taken into account in analyzing the total rates. The dominant $\nu_e - e$ scattering cross sections are decreased by about 2 % by radiative corrections.

The shape of the electron recoil spectrum produced by the scattering of $^8$B solar neutrinos is decreased by about 4 % at the highest electron energies. This effect should be included when analyzing future neutrino electron scattering experiments for the shape of the electron recoil energy spectrum.

If the shape of this spectrum can be measured accurately, the result will constitute a direct test of electroweak theory independent of solar models and will also constitute a way to discriminate among different proposed particle physics solutions that invoke physics beyond the standard electroweak model.

The analysis software for future experiments should take account of the fact that the photons produced by the radiative corrections (See the Appendix) are more strongly peaked in the forward direction than are the Cerenkov photons. In fact, the photons from radiative corrections will have an angular distribution similar to the well-known forward peaking of the electrons in neutrino-electron scattering (see for example [11]). The recoil energy of the electrons should be calculated primarily from the photons in the much wider Cerenkov cone (with appropriate corrections for incompleteness) in order not to confuse photons from the radiative corrections with photons from the Cerenkov light.

The shape of the electron recoil spectrum for incident $p - p$ and $^7$Be neutrinos is not appreciably affected by radiative corrections.

Radiative corrections will also affect the shape of the electron recoil spectrum produced when neutrinos are captured by nuclei. The effects analogous to those that we have calculated in this paper must, therefore, be evaluated for neutrinos captured by deuterium, the SNO experiment [5], and for neutrinos captured by argon, the ICARUS experiment [7].

# ACKNOWLEDGMENTS

J.N.B. was by supported by the NSF through Grant No. PHY-92-45317. M.K. was supported at the I.A.S. by the W. M. Keck Foundation and by the U.S. Department of Energy under contract DE-FG02-90ER40542, and at Columbia University under contract DE-FG02-92ER40699. A. Sirlin acknowledges support by the National Science Foundation under grant PHY-9313781. He would like to thank P. Gambino and M. Passera for carrying out the numerical checks described in Appendix B.

# APPENDIX A: RADIATIVE CORRECTIONS

We outline in this section the calculation of radiative corrections for neutrino-electron scattering: $\nu_\ell + e \to \nu_\ell + e \quad (\ell = e, \mu)$. The basic result for $\nu - e$ scattering is:

$$\frac{d\sigma}{dT} = \frac{2G_F^2 m}{\pi} \left\{ g_L^2(T) \left[ 1 + \frac{\alpha}{\pi} f_-(z) \right] \right.$$
$$\left. + g_R^2(T)(1-z)^2 \left[ 1 + \frac{\alpha}{\pi} f_+(z) \right] - g_R(T) g_L(T) \frac{m}{q} z \left[ 1 + \frac{\alpha}{\pi} f_{+-}(z) \right] \right\}, \quad (A1)$$

where $m$ is the electron mass, $T = E - m$ is the kinetic energy of recoil of the electron, $q$ is the incident neutrino energy, and $z = T/q$. We adopt $G_F = 1.16639 \times 10^{-5}$ GeV$^{-2}$ and $\sin^2 \hat{\theta}_W(m_Z) = 0.2317$ [19].

**For $\nu_e - e$ scattering:**

$$g_L^{(\nu_e,e)}(T) = \rho_{NC}^{(\nu,\ell)} \left[ \frac{1}{2} - \hat{\kappa}^{(\nu_e,e)}(T) \sin^2 \hat{\theta}_W(m_Z) \right] - 1 ,$$
$$g_R^{(\nu_e,e)}(T) = -\rho_{NC}^{(\nu,\ell)} \hat{\kappa}^{(\nu_e,e)}(T) \sin^2 \hat{\theta}_W(m_Z) , \quad (A2)$$

and

$$\rho_{NC}^{(\nu,\ell)} = 1.0126 \pm 0.0016 . \quad (A3)$$

The function $\hat{\kappa}^{(\nu_e,e)}$ is

$$\hat{\kappa}^{(\nu_e,e)}(T) = 0.9791 + 0.0097\, I(T) \pm 0.0025\,, \tag{A4}$$

where

$$I(T) \equiv \frac{1}{6}\left\{\frac{1}{3} + (3 - x^2)\left[\frac{1}{2}x \ln\left(\frac{x+1}{x-1}\right) - 1\right]\right\}\,, \tag{A5}$$

and $x = \sqrt{1 + \frac{2m}{T}}$. The deviations of $\rho_{NC}^{(\nu,\ell)}$ and $\hat{\kappa}^{(\nu_e,e)}(T)$ from 1 reflect the effect of the electroweak corrections. The $T$-dependence of $\hat{\kappa}$ comes from the diagrams shown in Fig. 7.

The error in $\rho_{NC}^{(\nu,\ell)}$ is mainly due to the lack of precise knowledge of $m_t$ and $m_H$. We use a top-quark mass $m_t = 177 \pm 11^{+18}_{-19}$ GeV [19]. The 177 GeV central value is for $m_H = 300$ GeV; the first error is the experimental uncertainty while the second reflects the shifts in the central value corresponding to $m_H = 1$ TeV, 60 GeV. The main error in $\hat{\kappa}^{(\nu_e,e)}(T)$ is of QCD origin and arises from diagrams such as that shown in Fig. 8.

We used a recent update of these QCD effects by Marciano [20].

**For $\nu_\mu - $ e scattering:**

$$g_L^{(\nu_\mu,e)}(T) = \rho_{NC}^{(\nu,\ell)}\left[\frac{1}{2} - \hat{\kappa}^{(\nu_\mu,e)}(T)\sin^2\hat{\theta}_W(m_Z)\right]\,,$$
$$g_R^{(\nu_\mu,e)}(T) = -\rho_{NC}^{(\nu,\ell)}\hat{\kappa}^{(\nu_\mu,e)}(T)\sin^2\hat{\theta}_W(m_Z)\,, \tag{A6}$$

Here $\rho_{NC}^{(\nu,\ell)}$ is the same as before but $\hat{\kappa}$ is changed:

$$\hat{\kappa}^{(\nu_\mu,e)}(T) = 0.9970 - 0.00037\, I(T) \pm 0.0025\,. \tag{A7}$$

The difference between $\hat{\kappa}^{(\nu_\mu,e)}(T)$ and $\hat{\kappa}^{(\nu_e,e)}(T)$ comes from the fact that the first diagram in Fig. 7 is replaced by the diagram in Fig. 9.

# APPENDIX B: QED EFFECTS

The functions $f_-(z)$, $f_+(z)$ and $f_{+-}(z)$ describe QED effects (virtual and real photons).

The functions $f_-(z)$ and $f_+(z)$ have been evaluated in the extreme relativistic approximation (E.R.) in Ref. [21]. This means that in that paper the electron mass $m$ has been set equal to zero in all contributions that remain finite as $m \to 0$. In practice, this means that the approximations $m/E \ll 1$ and $m/(E_{\max} - E) \ll 1$, $m/q \ll 1$ have been made. Exact expressions from which one can obtain $f_-(z)$ and $f_+(z)$ are given in Ref. [22]. The formulae are long and complicated. They can only be used by numerically tabulating them.

One of the main features of these functions is that there is a term that diverges logarithmically at the end point. As a consequence, $f_-(z)$ exhibits a sharp decrease (large negative values) near the end point. For $q \gg m$, this is not important for the $f_+(z)$ contributions because they are multiplied by $(1-z)^2$ which is very small near the end point. If $q \ll m$, $z_{\max} \ll 1$ and the above suppression is not present.

It turns out that the coefficient of this logarithm is exactly known, because it is related to the infrared divergence. It is universal, that is it appears in $f_-(z)$, $f_+(z)$ and $f_{+-}(z)$ and is given by

$$2\left[\frac{E}{\ell}\ln\left(\frac{E+\ell}{m}\right) - 1\right]\ln\left(1 - z - \frac{m}{E+\ell}\right), \tag{B1}$$

where $\ell = \sqrt{E^2 - m^2}$ is the three-momentum of the electron. At the end point,

$$z_{\max} = \frac{2q}{2q+m}, \quad \frac{m}{(E+\ell)_{\max}} = \frac{m}{2q+m}, \tag{B2}$$

and the argument of the logarithm vanishes.

In order to obtain a relatively simple formula that can be applied approximately in the non-relativistic domain, the ER expression for $f_-(z)$ has been modified to read

$$f_-(z) = \left[\frac{E}{\ell}\ln\left(\frac{E+\ell}{m}\right) - 1\right]\left[2\ln\left(1 - z - \frac{m}{E+\ell}\right) - \ln(1-z) - \frac{1}{2}\ln z - \frac{5}{12}\right]$$
$$+ \frac{1}{2}[L(z) - L(\beta)] - \frac{1}{2}\ln^2(1-z) - \left(\frac{11}{12} + \frac{z}{2}\right)\ln(1-z)$$
$$+ z\left[\ln z + \frac{1}{2}\ln\left(\frac{2q}{m}\right)\right] - \left(\frac{31}{18} + \frac{1}{12}\ln z\right)\beta - \frac{11}{12}z + \frac{z^2}{24}, \tag{B3}$$

where $L(x) = \int_0^x \ln|1-t|\frac{dt}{t}$ is the Spence function and $\beta = \frac{\ell}{E}$. A short table for $L(x)$ is given in a book by Lewin [23] about dilogarithms and related functions. The $L(x)$ used here is $L(x) = -Li_2(x)$, where $Li_2(x)$ is called the dilogarithm in Lewin's book, and is the function tabulated there. The function of $f_-(z)$ is plotted in Fig. 10 for neutrino energies of $q = 0.1$, 1, and 10 MeV.

In the E.R. approximation the formula for $f_-(z)$ reduces to that of Sarantakos et al. The modified formula contains exactly the log that blows up at the end point. This has the advantage that the error in the formula remains bounded no matter how close one approaches the end point [always assuming that we work to $\mathcal{O}(\alpha)$]. If one gets so close to the end point that $\frac{\alpha}{\pi}f_-(z) \approx -1$, perturbation theory breaks down and one has to consider multiple-photon emission. However, this is not a realistic possibility. The experimental resolution would have to be incredibly good before $\frac{\alpha}{\pi}f_-(z) \approx -1$. The function $f_-(z)$ has been constructed so that it vanishes at $z = 0$.*

The overall corrections can become sizeable, especially for large $q$, but this occurs only near the end point. That is mainly due to the above-mentioned logarithm. For small $q$, the coefficient of that log, namely $\left[\frac{E}{\ell}\ln\left(\frac{E+\ell}{m}\right) - 1\right]$, is small because $\ell \ll m$ and that factor vanishes as $\ell \to 0$.

In order to check the accuracy of Eq. (B3), that expression has been compared

---

*Note: the exact function $f_-(z)$ does not vanish at $z = 0$, but it gives a small correction there: $-(\alpha/\pi)f_-(0) = -\alpha/2\pi = -0.12\%$.

numerically with the corresponding result from Ref. [22], for $q = 0.2$ MeV, 1 MeV, 5 MeV, and 15 MeV, and several values of $T$. As expected, in the non-relativistic regime the relative error of Eq. (B3) is rather large. However, this occurs when the correction itself is quite small. As a consequence, in all the cases considered, the absolute value of the error remains small, typically $\sim 0.1\%$. For 5 MeV $\lesssim q \lesssim$ 15 MeV, the absolute error reaches $\approx -0.4\%$, but this only occurs very near the end-point, where the correction itself is very sizable.

The corresponding modification of the ER expression for $(1-z)^2 f_+(z)$ is

$$
\begin{aligned}
(1-z)^2 f_+(z) = & \left[\frac{E}{\ell} \ln\left(\frac{E+\ell}{m}\right) - 1\right] \\
& \times \left\{(1-z)^2 \left[2\ln\left(1 - z - \frac{m}{E+\ell}\right) - \ln(1-z) - \frac{\ln z}{2} - \frac{2}{3}\right]\right. \\
& \left. - \frac{(z^2 \ln z + 1 - z)}{2}\right\} \\
& - \frac{(1-z)^2}{2}\left\{\ln^2(1-z) + \beta\left[L(1-z) - \ln z \ln(1-z)\right]\right\} \\
& + \ln(1-z)\left[\frac{z^2}{2}\ln z + \frac{(1-z)}{3}\left(2z - \frac{1}{2}\right)\right] \\
& - \frac{z^2}{2}L(1-z) - \frac{z(1-2z)}{3}\ln z - \frac{z(1-z)}{6} \\
& - \frac{\beta}{12}\left[\ln z + (1-z)\left(\frac{115 - 109z}{6}\right)\right]
\end{aligned} \tag{B4}
$$

This formula reduces in the E.R. limit to that of Sarantakos et al., is 0 at $\ell = 0$, contains correctly the logarithm discussed before and, for $z \to 1$, behaves as $(1-z)^2$ modulo logs, which is the correct behavior. The function $f_+(z)$ is plotted in Fig. 11 for neutrino energies of $q = 0.1$, 1, and 10 MeV.

The function $f_{+-}(z)$ has not been calculated previously. The maximum value of $mz/q$ in the last term of $d\sigma/dT$ is $2m/(2q+m)$. So there is a suppression factor for $q \gg m$. For $q \lesssim m$, there is no suppression from this factor. If $q \ll m$, the electron is non-relativistic and the corrections are expected to be very small, except very close to the end point (because of the divergent logarithm). The largest contributions from $f_{+-}(z)$ should occur when $q \approx m$ and one is close to the end point. For $\nu_e - e$ there is another suppression factor because $|g_R|$ is substantially smaller than $|g_L|$. The approximation adopted here is to set

$$f_{+-}(z) = \left[\frac{E}{\ell}\ln\left(\frac{E+\ell}{m}\right) - 1\right] 2\ln\left(1 - z - \frac{m}{E+\ell}\right) . \tag{B5}$$

This should take care of the largest effects near the end point. Aside from this, the other contributions from $f_{+-}(z)$ to $d\sigma/dT$ should be small. The function $f_{+-}(z)$ is plotted in Fig. 12 for neutrino energies of $q = 0.1$, 1, and 10 MeV.

FIGURE CAPTIONS

FIG. 1. Electron recoil spectrum from the 0.862 MeV $^7$Be neutrino line. The heavy (light) solid curve is the ratio—minus unity—of the probability distribution for $\nu_e$ ($\nu_\mu$ or $\nu_\tau$) scattering obtained using an updated value of $\sin^2 \hat{\theta}_W(m_Z) = 0.2317$ and including radiative corrections to the probability distribution obtained using the less accurate value 0.23 with no radiative corrections. The plotted ratio gives the net change, less than 1% over essentially all of the allowed recoil-energy range, in the probability distribution that is obtained using the formulae and constants employed in this paper compared to the values that were obtained in refs. [11,15]. The effect of changing just $\sin^2 \hat{\theta}_W$ is shown in the dotted curves. The dotted curves are the ratios—minus unity—of the probability distributions that are obtained using $\sin^2 \hat{\theta}_W = 0.2317$ to the probability distributions obtained using the older value of $\sin^2 \hat{\theta}_W = 0.23$. The dashed curves show the effects of just including radiative corrections. The dashed curves are the ratios—minus unity—of the probability distributions obtained including radiative corrections to those obtained without radiative corrections. The spectra are displayed up to $T = 0.99 T_{\max}$.

FIG. 2. Same as Fig. 1 but for the $^7$Be neutrino line with an energy of 0.384 MeV.

FIG. 3. Same as Fig. 1 for the *pp* neutrinos.

FIG. 4. Same as Fig. 1, except that the dotted curves are omitted, for the $^8$B neutrinos.

FIG. 5. Recoil spectra for both electron and muon neutrinos from $p-p$, $^7$Be, $^{13}$N, and $^{15}$O neutrinos as a function of electron kinetic energy, $T$. The spectra are normalized such that the total probability is unity. The dotted curve is the recoil spectrum from $^{13}$N neutrinos and the short-dash curve is that from $^{15}$O neutrinos. The solid curve is the recoil spectrum from the 0.862-MeV $^7$Be line, the dot-dash curve is that from the 0.384-MeV $^7$Be line, and the long-dash curve is that from the $p-p$ neutrinos.

FIG. 6. Ratios of the total cross section for $\nu-e$ scattering as a function of the neutrino energy $q$. The meanings of the different symbols for the curves are the same as in Fig. 1.

FIG. 7. Feynman diagrams for electroweak corrections to $\nu_e - e$ scattering.

FIG. 8. Feynman diagram for QCD corrections to $\kappa^{(\nu_e, e)}(T)$.

FIG. 9. One of the Feynman diagrams for $\nu_\mu - e$ scattering.

FIG. 10. The function $f_-(z)$ as a function of $z$ for neutrino energies of $q = 0.1$ MeV (solid curve), $q = 1$ MeV (short-dash curve), and $q = 10$ MeV (long-dash curve).

FIG. 11. The function $f_+(z)$ as a function of $z$ for neutrino energies of $q = 0.1$ MeV (solid curve), $q = 1$ MeV (short-dash curve), and $q = 10$ MeV (long-dash curve).

FIG. 12. The function $f_\pm(z)$ as a function of $z$ for neutrino energies of $q = 0.1$ MeV (solid curve), $q = 1$ MeV (short-dash curve), and $q = 10$ MeV (long-dash curve).

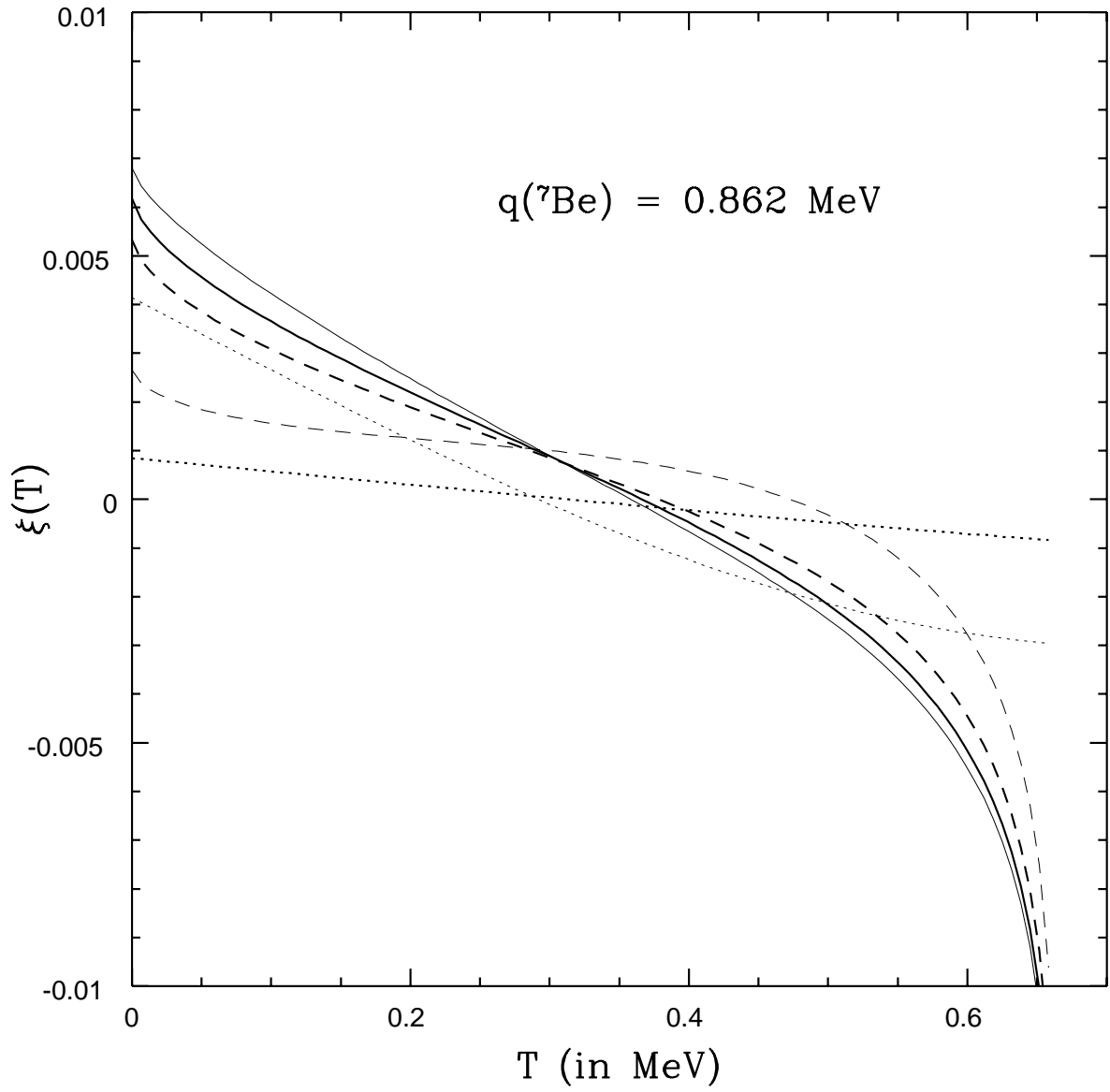

FIG. 1

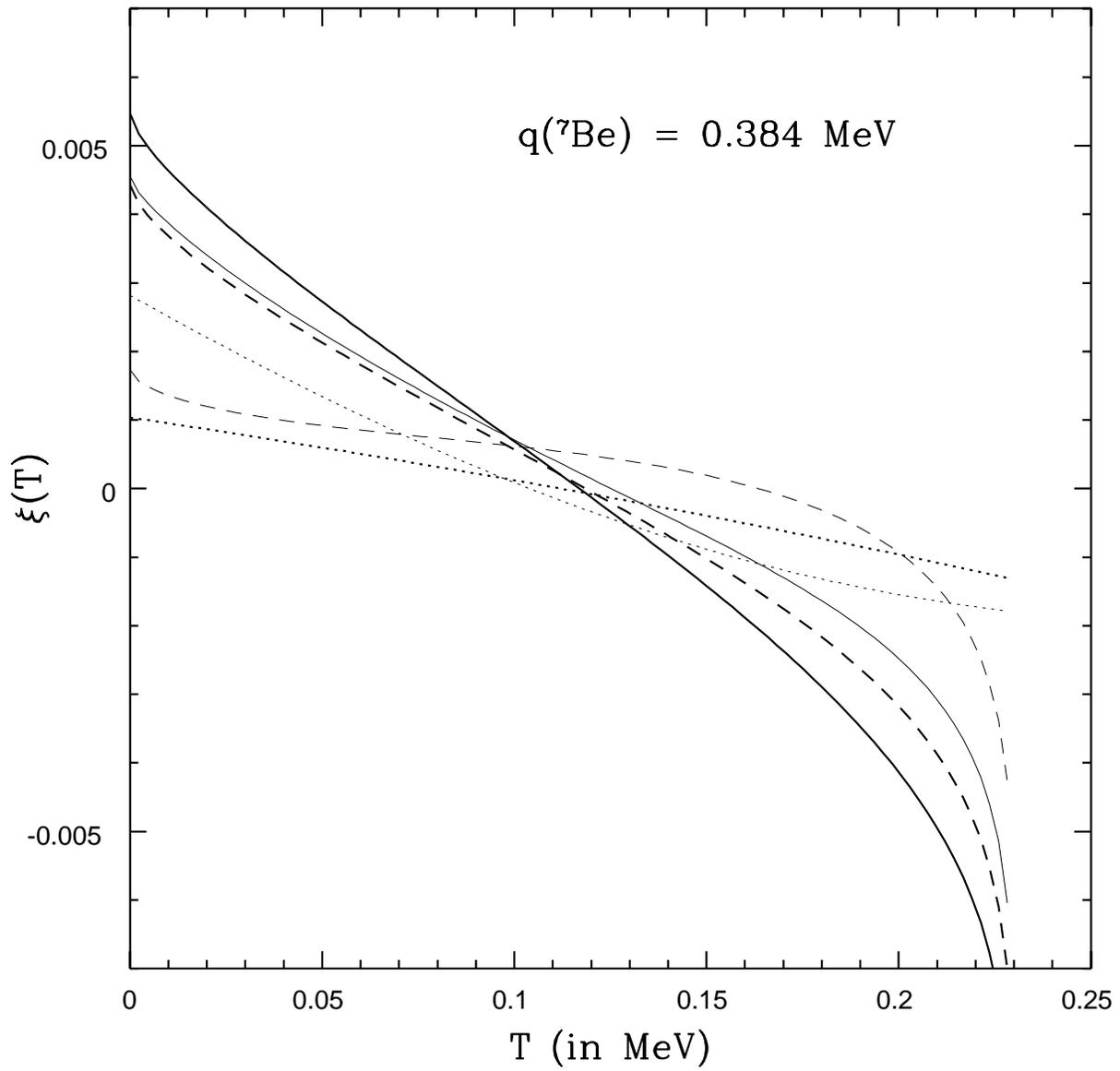

FIG. 2

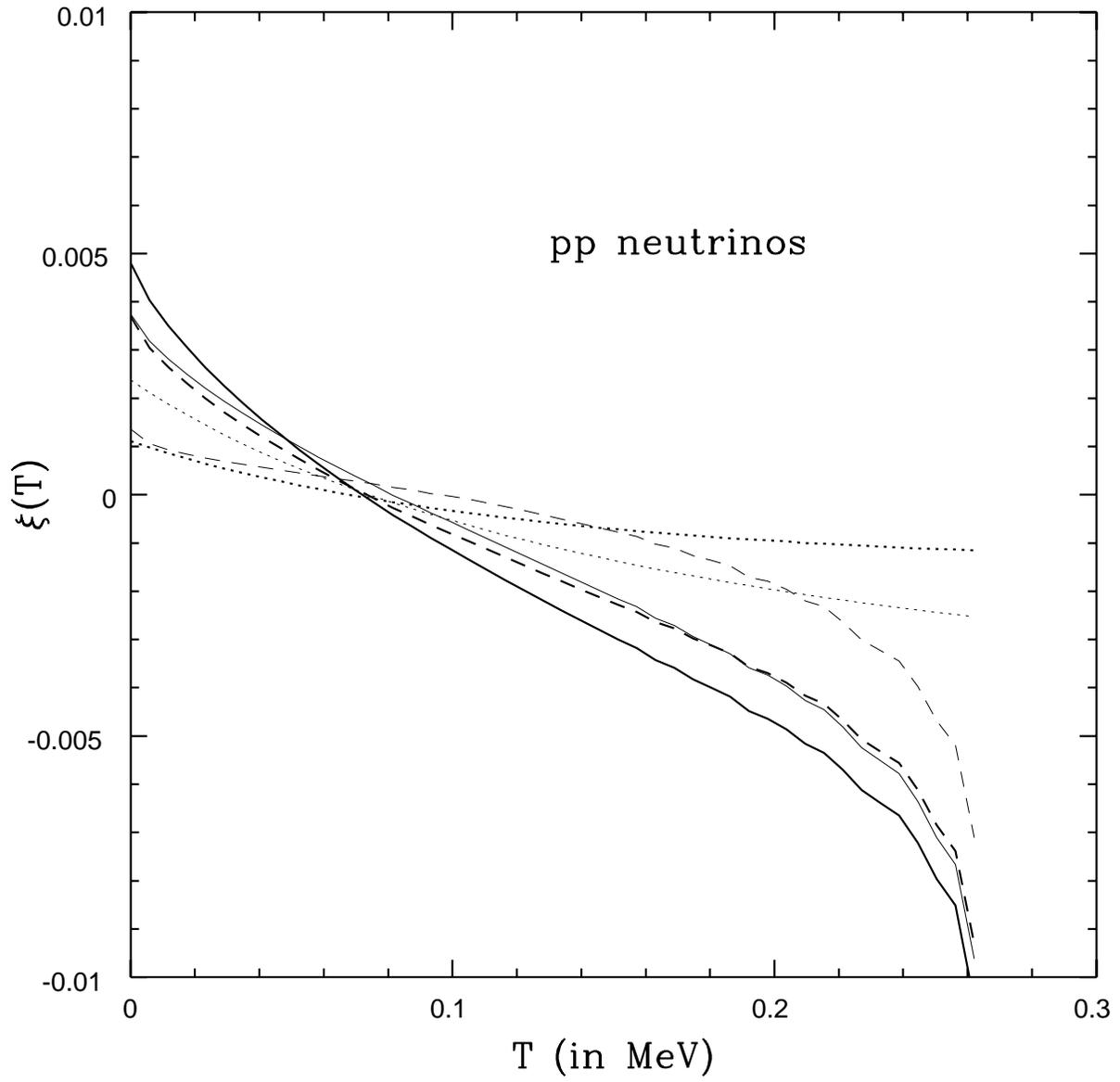

FIG. 3

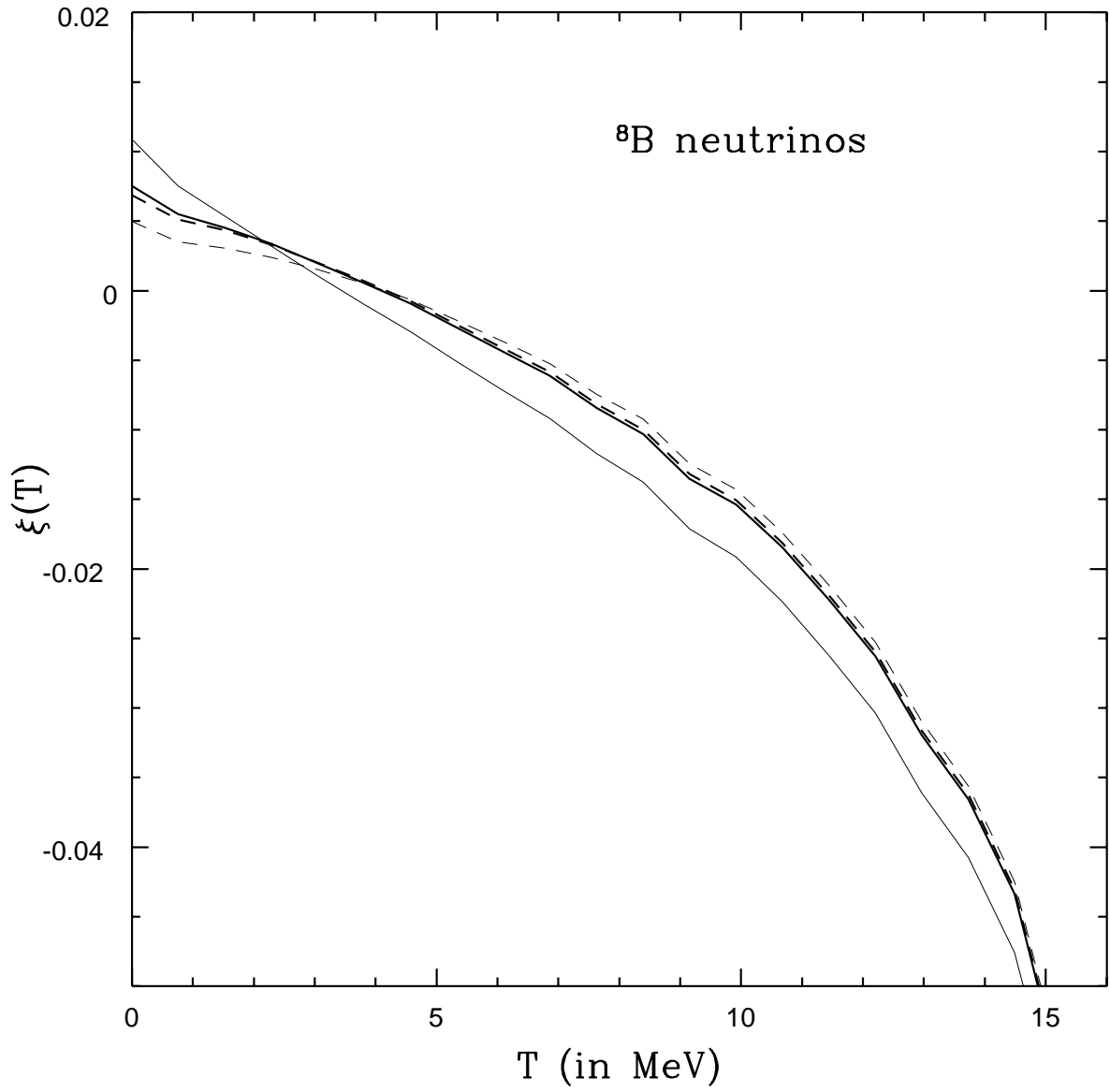

FIG. 4

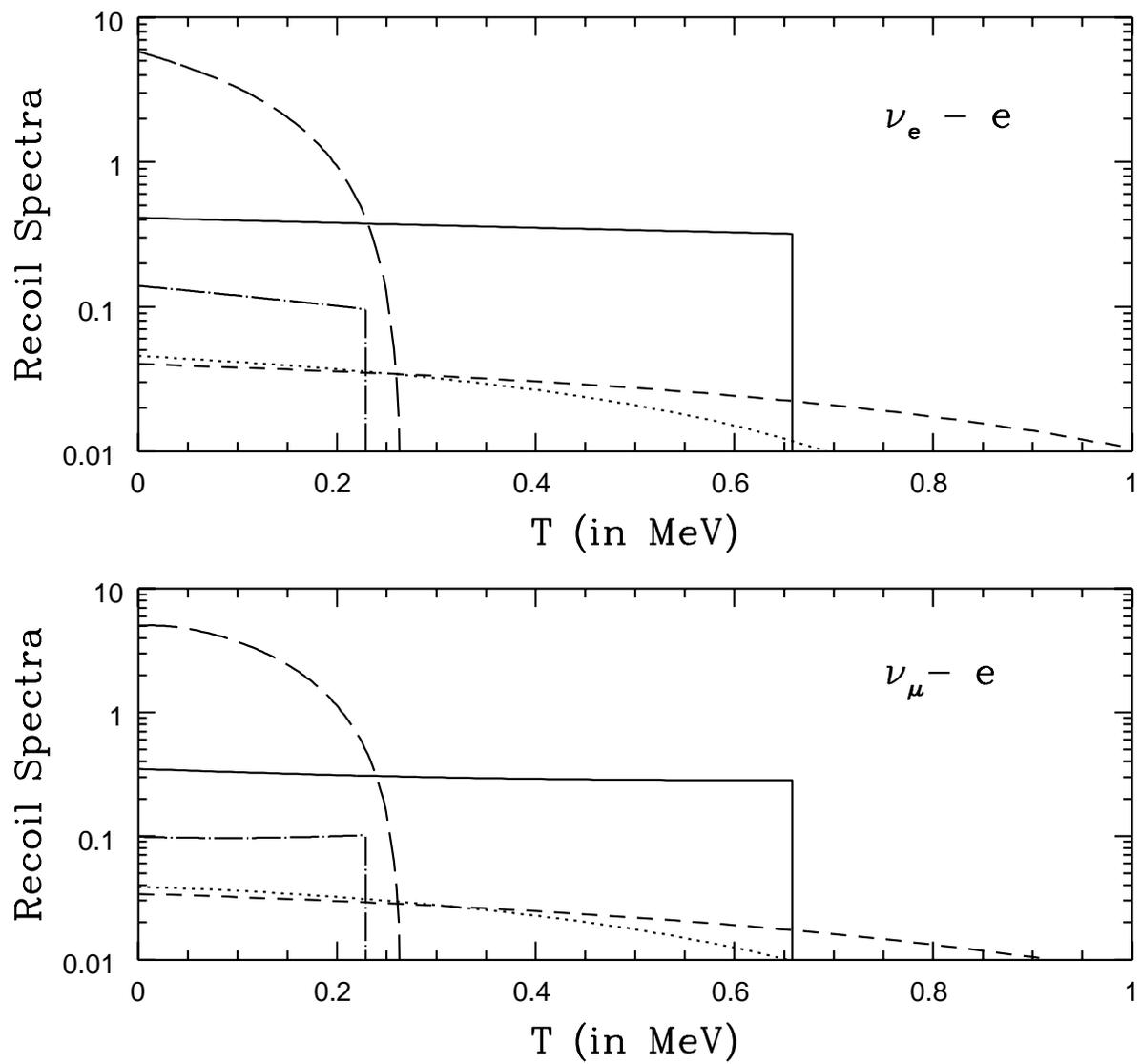

FIG. 5

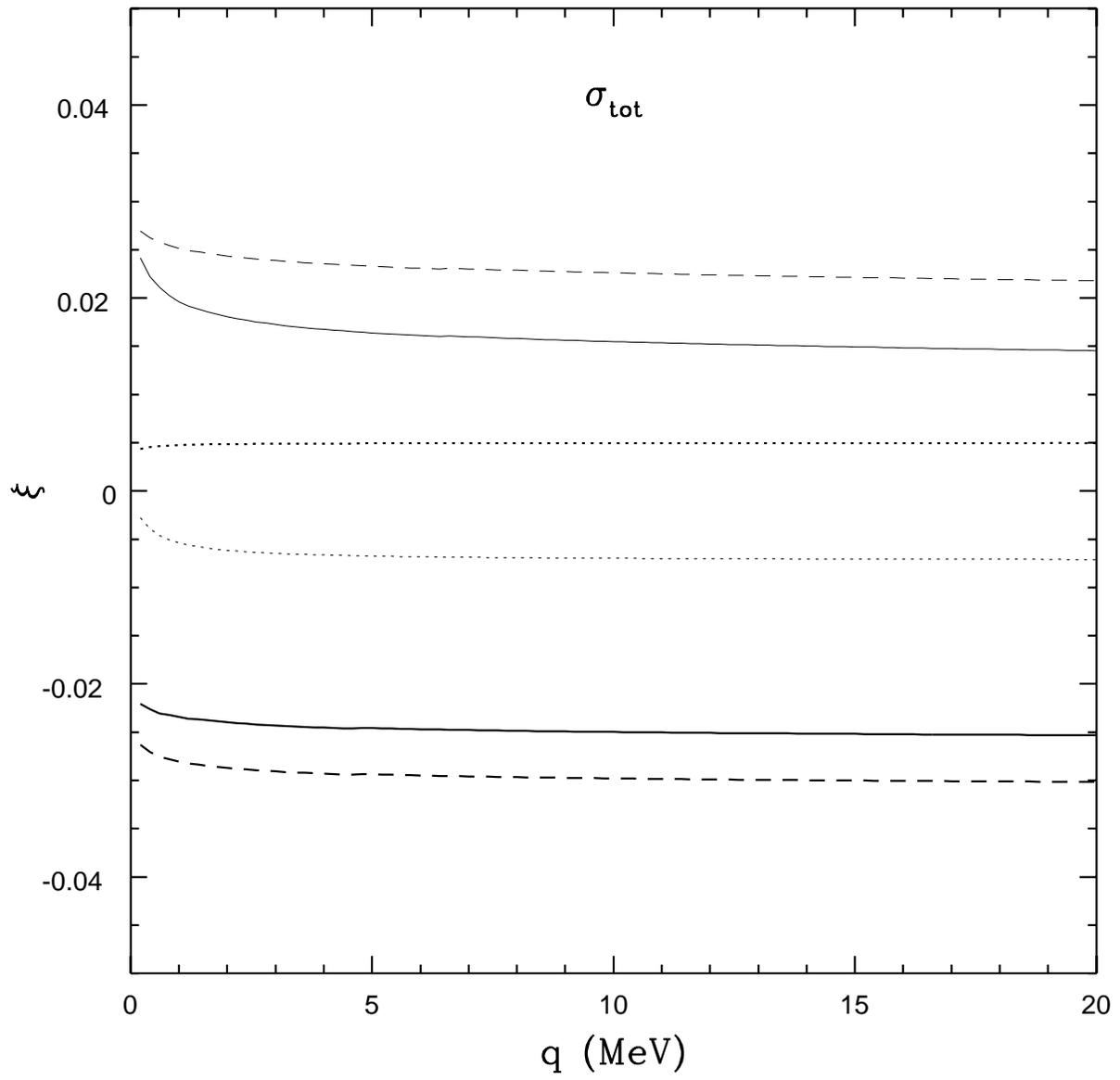

FIG. 6

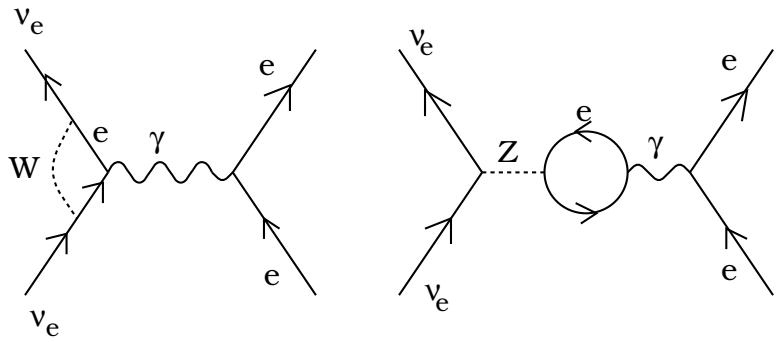

FIG. 7

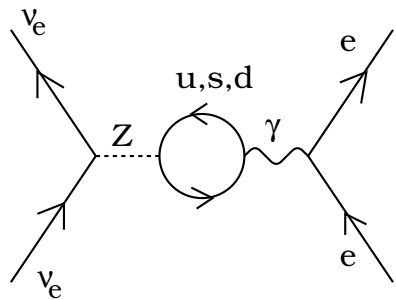

FIG. 8

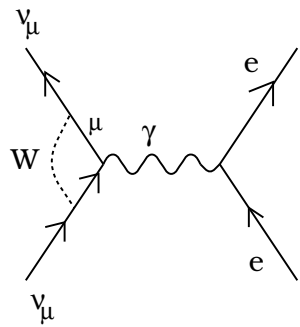

FIG. 9

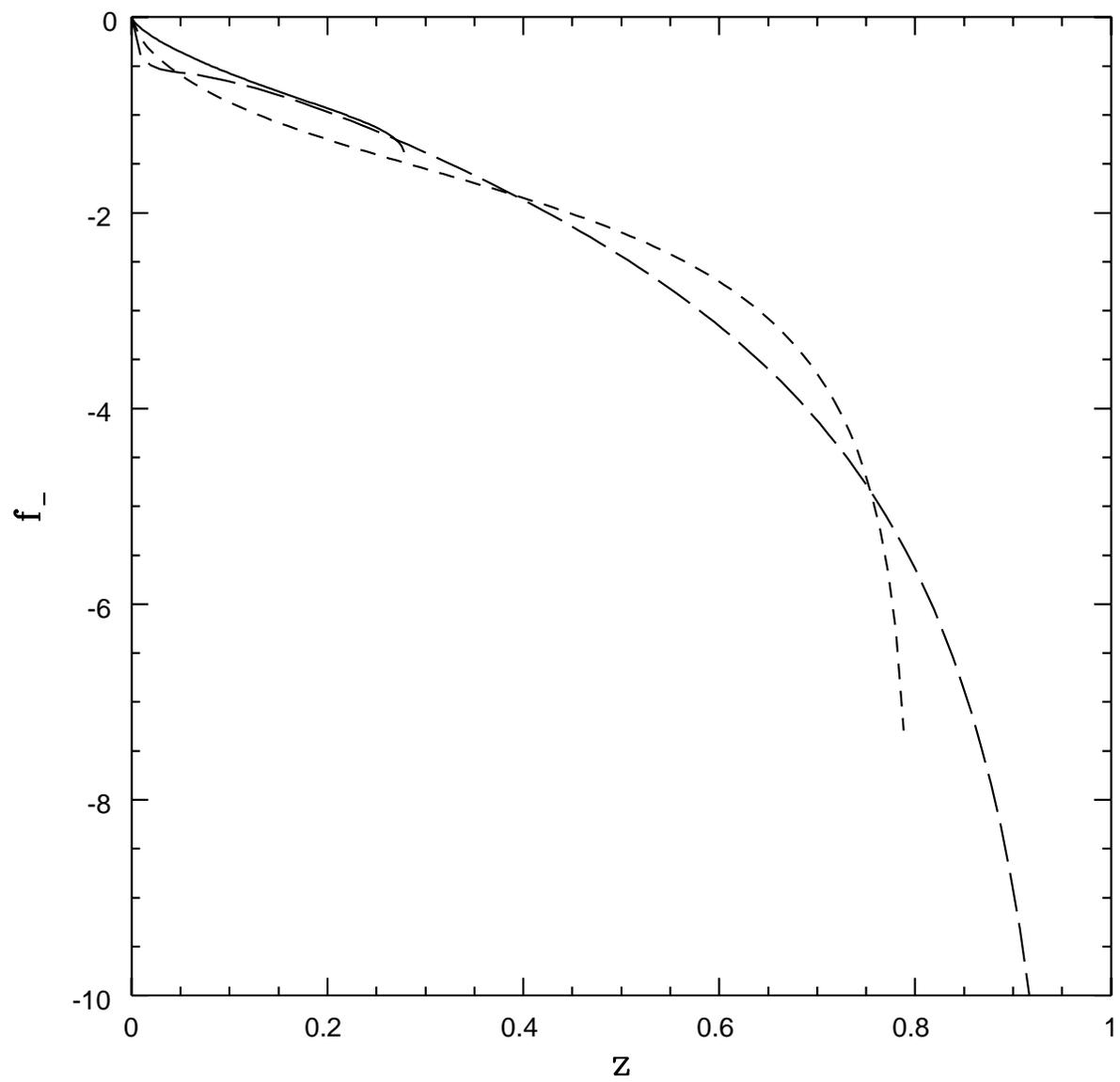

FIG. 10

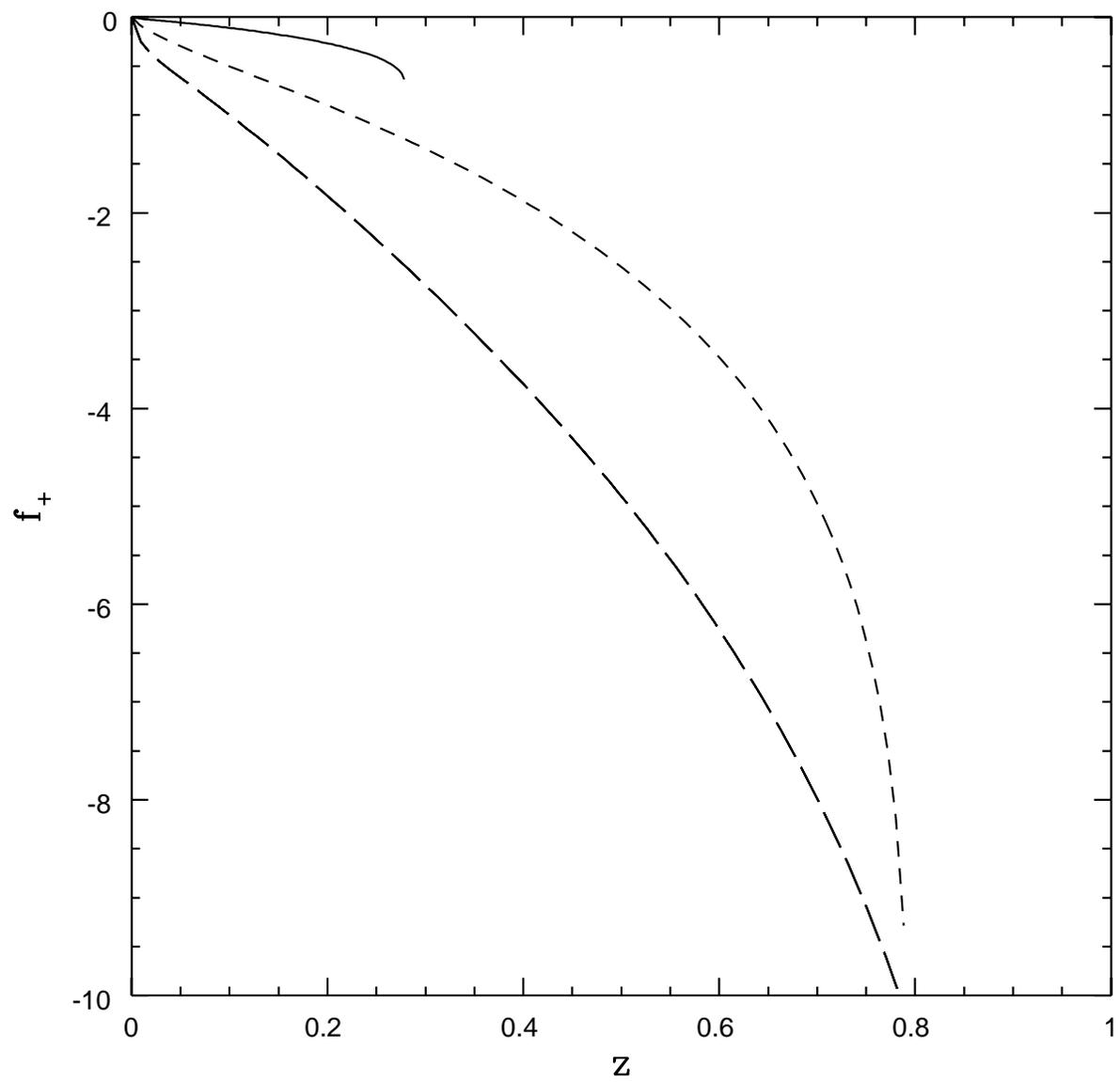

FIG. 11

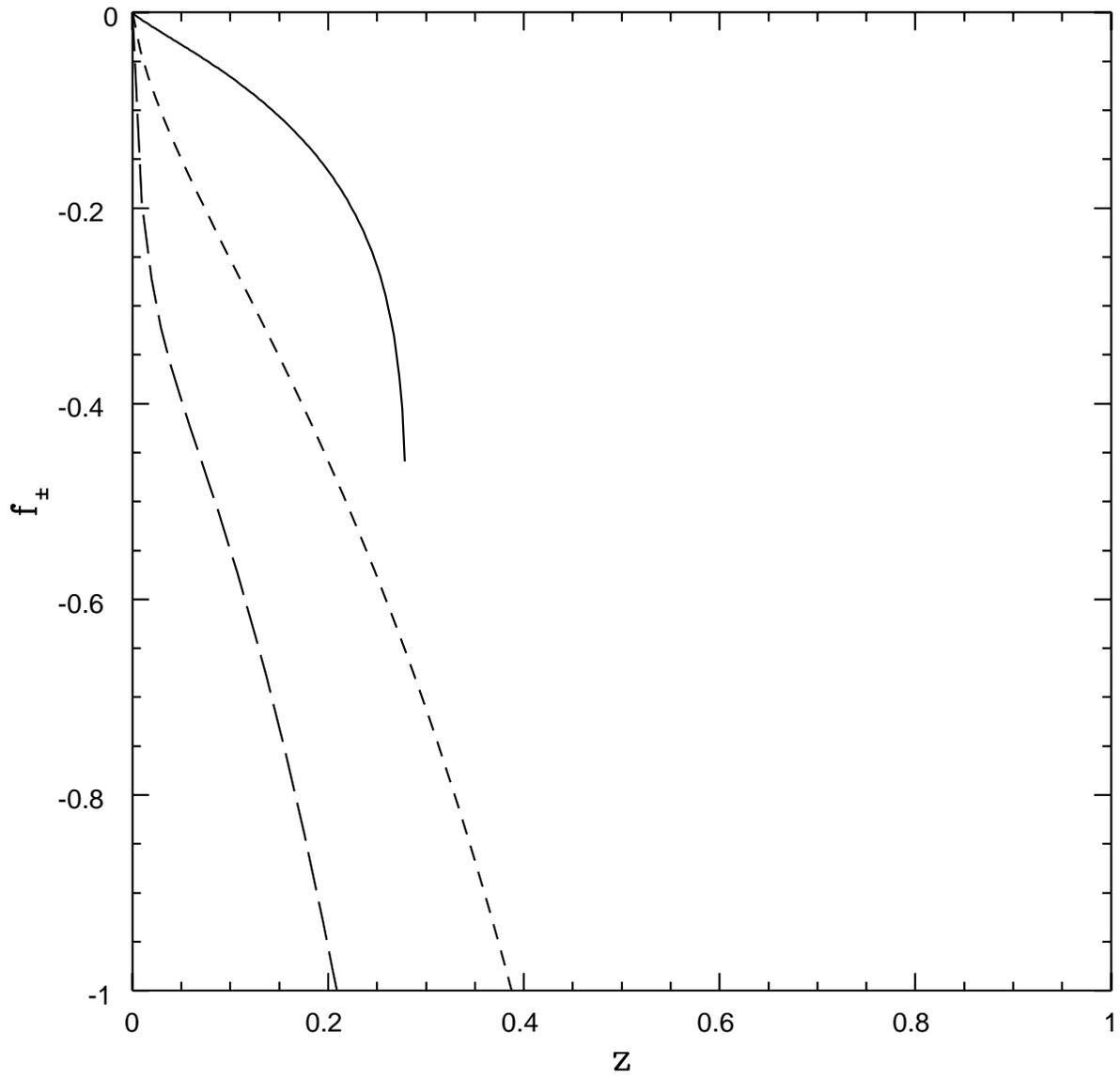

FIG. 12

TABLE I. Recoil spectrum from $\nu_e - e$ and $\nu_\mu - e$ scattering by the $^7$Be neutrino line with energy 0.862 MeV. The kinetic energy of the electron is denoted by $T$ (measured in MeV) and the normalized probability distributions per MeV by $P(T)_{\nu_e}$ and $P(T)_{\nu_\mu}$.

| $T$ | $P(T)_{\nu_e}$ | $P(T)_{\nu_\mu}$ | $T$ | $P(T)_{\nu_e}$ | $P(T)_{\nu_\mu}$ |
|---|---|---|---|---|---|
| 0.0000 | 1.7143 | 1.7367 | 0.3325 | 1.4970 | 1.4686 |
| 0.0066 | 1.7089 | 1.7287 | 0.3391 | 1.4932 | 1.4655 |
| 0.0133 | 1.7039 | 1.7211 | 0.3458 | 1.4894 | 1.4625 |
| 0.0199 | 1.6990 | 1.7136 | 0.3524 | 1.4856 | 1.4596 |
| 0.0266 | 1.6941 | 1.7063 | 0.3591 | 1.4818 | 1.4567 |
| 0.0332 | 1.6893 | 1.6990 | 0.3657 | 1.4781 | 1.4540 |
| 0.0399 | 1.6845 | 1.6919 | 0.3724 | 1.4744 | 1.4513 |
| 0.0465 | 1.6797 | 1.6849 | 0.3790 | 1.4707 | 1.4488 |
| 0.0532 | 1.6750 | 1.6780 | 0.3857 | 1.4670 | 1.4463 |
| 0.0598 | 1.6703 | 1.6711 | 0.3923 | 1.4633 | 1.4439 |
| 0.0665 | 1.6656 | 1.6644 | 0.3990 | 1.4596 | 1.4416 |
| 0.0731 | 1.6610 | 1.6577 | 0.4056 | 1.4560 | 1.4393 |
| 0.0798 | 1.6564 | 1.6512 | 0.4123 | 1.4524 | 1.4372 |
| 0.0864 | 1.6518 | 1.6447 | 0.4189 | 1.4488 | 1.4351 |
| 0.0931 | 1.6472 | 1.6383 | 0.4255 | 1.4452 | 1.4332 |
| 0.0997 | 1.6427 | 1.6321 | 0.4322 | 1.4416 | 1.4313 |
| 0.1064 | 1.6382 | 1.6259 | 0.4388 | 1.4381 | 1.4295 |
| 0.1130 | 1.6337 | 1.6198 | 0.4455 | 1.4345 | 1.4277 |
| 0.1197 | 1.6292 | 1.6138 | 0.4521 | 1.4310 | 1.4261 |
| 0.1263 | 1.6248 | 1.6079 | 0.4588 | 1.4275 | 1.4245 |
| 0.1330 | 1.6204 | 1.6020 | 0.4654 | 1.4240 | 1.4231 |
| 0.1396 | 1.6160 | 1.5963 | 0.4721 | 1.4205 | 1.4217 |
| 0.1463 | 1.6116 | 1.5906 | 0.4787 | 1.4171 | 1.4204 |
| 0.1529 | 1.6072 | 1.5851 | 0.4854 | 1.4136 | 1.4192 |
| 0.1596 | 1.6029 | 1.5796 | 0.4920 | 1.4102 | 1.4180 |
| 0.1662 | 1.5986 | 1.5742 | 0.4987 | 1.4067 | 1.4170 |
| 0.1729 | 1.5943 | 1.5690 | 0.5053 | 1.4033 | 1.4160 |
| 0.1795 | 1.5900 | 1.5638 | 0.5120 | 1.4000 | 1.4151 |
| 0.1862 | 1.5857 | 1.5587 | 0.5186 | 1.3966 | 1.4143 |
| 0.1928 | 1.5815 | 1.5536 | 0.5253 | 1.3932 | 1.4135 |
| 0.1995 | 1.5773 | 1.5487 | 0.5319 | 1.3898 | 1.4129 |
| 0.2061 | 1.5731 | 1.5439 | 0.5386 | 1.3865 | 1.4123 |
| 0.2128 | 1.5689 | 1.5391 | 0.5452 | 1.3831 | 1.4118 |
| 0.2194 | 1.5647 | 1.5344 | 0.5519 | 1.3798 | 1.4113 |
| 0.2261 | 1.5606 | 1.5299 | 0.5585 | 1.3764 | 1.4110 |
| 0.2327 | 1.5565 | 1.5254 | 0.5652 | 1.3731 | 1.4107 |
| 0.2394 | 1.5524 | 1.5210 | 0.5718 | 1.3698 | 1.4104 |
| 0.2460 | 1.5483 | 1.5167 | 0.5785 | 1.3664 | 1.4103 |
| 0.2527 | 1.5442 | 1.5124 | 0.5851 | 1.3631 | 1.4102 |
| 0.2593 | 1.5402 | 1.5083 | 0.5918 | 1.3598 | 1.4101 |
| 0.2660 | 1.5362 | 1.5043 | 0.5984 | 1.3564 | 1.4101 |
| 0.2726 | 1.5322 | 1.5003 | 0.6051 | 1.3530 | 1.4101 |
| 0.2793 | 1.5282 | 1.4964 | 0.6117 | 1.3496 | 1.4102 |
| 0.2859 | 1.5242 | 1.4926 | 0.6184 | 1.3461 | 1.4103 |
| 0.2926 | 1.5203 | 1.4889 | 0.6250 | 1.3426 | 1.4104 |
| 0.2992 | 1.5164 | 1.4853 | 0.6317 | 1.3390 | 1.4104 |
| 0.3059 | 1.5125 | 1.4818 | 0.6383 | 1.3352 | 1.4104 |
| 0.3125 | 1.5086 | 1.4784 | 0.6450 | 1.3312 | 1.4101 |
| 0.3192 | 1.5047 | 1.4750 | 0.6516 | 1.3267 | 1.4095 |
| 0.3258 | 1.5008 | 1.4717 | 0.6583 | 1.3211 | 1.4077 |

TABLE II. Recoil spectrum from $\nu_e - e$ and $\nu_\mu - e$ scattering by the $^7$Be neutrino line with energy 0.384 MeV. The kinetic energy of the electron is denoted by $T$ (measured in MeV) and the normalized probability distributions per MeV by $P(T)_{\nu_e}$ and $P(T)_{\nu_\mu}$.

| $T$ | $P(T)_{\nu_e}$ | $P(T)_{\nu_\mu}$ | $T$ | $P(T)_{\nu_e}$ | $P(T)_{\nu_\mu}$ |
|---|---|---|---|---|---|
| 0.0000 | 5.1781 | 4.3774 | 0.1153 | 4.3232 | 4.2830 |
| 0.0023 | 5.1591 | 4.3716 | 0.1176 | 4.3071 | 4.2846 |
| 0.0046 | 5.1407 | 4.3663 | 0.1199 | 4.2910 | 4.2863 |
| 0.0069 | 5.1226 | 4.3612 | 0.1222 | 4.2750 | 4.2881 |
| 0.0092 | 5.1045 | 4.3563 | 0.1245 | 4.2590 | 4.2901 |
| 0.0115 | 5.0865 | 4.3515 | 0.1268 | 4.2430 | 4.2922 |
| 0.0138 | 5.0686 | 4.3470 | 0.1291 | 4.2271 | 4.2945 |
| 0.0161 | 5.0507 | 4.3425 | 0.1314 | 4.2112 | 4.2968 |
| 0.0184 | 5.0329 | 4.3382 | 0.1337 | 4.1954 | 4.2993 |
| 0.0208 | 5.0151 | 4.3341 | 0.1360 | 4.1795 | 4.3020 |
| 0.0231 | 4.9974 | 4.3301 | 0.1384 | 4.1638 | 4.3047 |
| 0.0254 | 4.9798 | 4.3262 | 0.1407 | 4.1480 | 4.3076 |
| 0.0277 | 4.9622 | 4.3225 | 0.1430 | 4.1323 | 4.3106 |
| 0.0300 | 4.9446 | 4.3190 | 0.1453 | 4.1166 | 4.3138 |
| 0.0323 | 4.9271 | 4.3155 | 0.1476 | 4.1010 | 4.3171 |
| 0.0346 | 4.9097 | 4.3122 | 0.1499 | 4.0853 | 4.3205 |
| 0.0369 | 4.8922 | 4.3091 | 0.1522 | 4.0697 | 4.3241 |
| 0.0392 | 4.8749 | 4.3061 | 0.1545 | 4.0542 | 4.3277 |
| 0.0415 | 4.8575 | 4.3032 | 0.1568 | 4.0387 | 4.3315 |
| 0.0438 | 4.8402 | 4.3005 | 0.1591 | 4.0232 | 4.3355 |
| 0.0461 | 4.8230 | 4.2979 | 0.1614 | 4.0077 | 4.3395 |
| 0.0484 | 4.8057 | 4.2954 | 0.1637 | 3.9923 | 4.3437 |
| 0.0507 | 4.7886 | 4.2931 | 0.1660 | 3.9769 | 4.3480 |
| 0.0530 | 4.7714 | 4.2909 | 0.1683 | 3.9615 | 4.3525 |
| 0.0553 | 4.7544 | 4.2889 | 0.1706 | 3.9462 | 4.3570 |
| 0.0576 | 4.7373 | 4.2870 | 0.1729 | 3.9309 | 4.3617 |
| 0.0600 | 4.7203 | 4.2852 | 0.1752 | 3.9156 | 4.3666 |
| 0.0623 | 4.7033 | 4.2835 | 0.1775 | 3.9003 | 4.3715 |
| 0.0646 | 4.6864 | 4.2820 | 0.1799 | 3.8851 | 4.3765 |
| 0.0669 | 4.6695 | 4.2807 | 0.1822 | 3.8699 | 4.3817 |
| 0.0692 | 4.6526 | 4.2794 | 0.1845 | 3.8547 | 4.3870 |
| 0.0715 | 4.6358 | 4.2783 | 0.1868 | 3.8396 | 4.3924 |
| 0.0738 | 4.6190 | 4.2774 | 0.1891 | 3.8245 | 4.3980 |
| 0.0761 | 4.6023 | 4.2765 | 0.1914 | 3.8093 | 4.4036 |
| 0.0784 | 4.5856 | 4.2758 | 0.1937 | 3.7943 | 4.4094 |
| 0.0807 | 4.5689 | 4.2753 | 0.1960 | 3.7792 | 4.4153 |
| 0.0830 | 4.5522 | 4.2749 | 0.1983 | 3.7641 | 4.4212 |
| 0.0853 | 4.5357 | 4.2746 | 0.2006 | 3.7491 | 4.4273 |
| 0.0876 | 4.5191 | 4.2744 | 0.2029 | 3.7341 | 4.4335 |
| 0.0899 | 4.5026 | 4.2744 | 0.2052 | 3.7191 | 4.4398 |
| 0.0922 | 4.4861 | 4.2745 | 0.2075 | 3.7040 | 4.4461 |
| 0.0945 | 4.4696 | 4.2748 | 0.2098 | 3.6890 | 4.4525 |
| 0.0968 | 4.4532 | 4.2751 | 0.2121 | 3.6739 | 4.4590 |
| 0.0992 | 4.4368 | 4.2756 | 0.2144 | 3.6588 | 4.4655 |
| 0.1015 | 4.4205 | 4.2763 | 0.2167 | 3.6437 | 4.4720 |
| 0.1038 | 4.4042 | 4.2771 | 0.2191 | 3.6284 | 4.4785 |
| 0.1061 | 4.3879 | 4.2780 | 0.2214 | 3.6130 | 4.4849 |
| 0.1084 | 4.3717 | 4.2790 | 0.2237 | 3.5974 | 4.4911 |
| 0.1107 | 4.3555 | 4.2802 | 0.2260 | 3.5812 | 4.4966 |
| 0.1130 | 4.3393 | 4.2815 | 0.2283 | 3.5638 | 4.5007 |

TABLE III. Recoil spectrum from $\nu_e - e$ and $\nu_\mu - e$ scattering by $pp$ neutrinos. The kinetic energy of the electron is denoted by $T$ (measured in MeV) and the normalized probability distributions per MeV by $P(T)_{\nu_e}$ and $P(T)_{\nu_\mu}$.

| $T$ | $P(T)_{\nu_e}$ | $P(T)_{\nu_\mu}$ | $T$ | $P(T)_{\nu_e}$ | $P(T)_{\nu_\mu}$ |
|---|---|---|---|---|---|
| 0.0000 | 8.5321 | 6.7549 | 0.1326 | 3.5989 | 3.9363 |
| 0.0027 | 8.4161 | 6.8009 | 0.1353 | 3.5115 | 3.8500 |
| 0.0053 | 8.3053 | 6.8314 | 0.1379 | 3.4097 | 3.7422 |
| 0.0080 | 8.1969 | 6.8504 | 0.1406 | 3.3227 | 3.6551 |
| 0.0106 | 8.0902 | 6.8610 | 0.1432 | 3.2212 | 3.5464 |
| 0.0133 | 7.9855 | 6.8659 | 0.1459 | 3.1347 | 3.4585 |
| 0.0159 | 7.8815 | 6.8631 | 0.1485 | 3.0334 | 3.3490 |
| 0.0186 | 7.7779 | 6.8530 | 0.1512 | 2.9475 | 3.2606 |
| 0.0212 | 7.6766 | 6.8415 | 0.1538 | 2.8621 | 3.1721 |
| 0.0239 | 7.5733 | 6.8182 | 0.1565 | 2.7614 | 3.0620 |
| 0.0265 | 7.4724 | 6.7953 | 0.1592 | 2.6768 | 2.9735 |
| 0.0292 | 7.3748 | 6.7746 | 0.1618 | 2.5925 | 2.8847 |
| 0.0318 | 7.2744 | 6.7421 | 0.1645 | 2.4930 | 2.7748 |
| 0.0345 | 7.1738 | 6.7047 | 0.1671 | 2.4097 | 2.6862 |
| 0.0371 | 7.0771 | 6.6718 | 0.1698 | 2.3110 | 2.5766 |
| 0.0398 | 6.9762 | 6.6257 | 0.1724 | 2.2287 | 2.4885 |
| 0.0424 | 6.8797 | 6.5854 | 0.1751 | 2.1471 | 2.4009 |
| 0.0451 | 6.7782 | 6.5310 | 0.1777 | 2.0501 | 2.2923 |
| 0.0477 | 6.6817 | 6.4837 | 0.1804 | 1.9697 | 2.2053 |
| 0.0504 | 6.5851 | 6.4333 | 0.1830 | 1.8900 | 2.1187 |
| 0.0531 | 6.4886 | 6.3799 | 0.1857 | 1.7950 | 2.0119 |
| 0.0557 | 6.3920 | 6.3236 | 0.1883 | 1.7168 | 1.9265 |
| 0.0584 | 6.2954 | 6.2645 | 0.1910 | 1.6396 | 1.8419 |
| 0.0610 | 6.1985 | 6.2026 | 0.1936 | 1.5630 | 1.7578 |
| 0.0637 | 6.1015 | 6.1379 | 0.1963 | 1.4716 | 1.6544 |
| 0.0663 | 6.0042 | 6.0707 | 0.1989 | 1.3970 | 1.5721 |
| 0.0690 | 5.9068 | 6.0010 | 0.2016 | 1.3233 | 1.4906 |
| 0.0716 | 5.8091 | 5.9289 | 0.2042 | 1.2353 | 1.3908 |
| 0.0743 | 5.7111 | 5.8543 | 0.2069 | 1.1640 | 1.3116 |
| 0.0769 | 5.6128 | 5.7775 | 0.2095 | 1.0937 | 1.2334 |
| 0.0796 | 5.5142 | 5.6984 | 0.2122 | 1.0245 | 1.1564 |
| 0.0822 | 5.4250 | 5.6338 | 0.2149 | 0.9419 | 1.0625 |
| 0.0849 | 5.3262 | 5.5510 | 0.2175 | 0.8756 | 0.9884 |
| 0.0875 | 5.2270 | 5.4661 | 0.2202 | 0.8106 | 0.9157 |
| 0.0902 | 5.1275 | 5.3792 | 0.2228 | 0.7470 | 0.8445 |
| 0.0928 | 5.0385 | 5.3084 | 0.2255 | 0.6850 | 0.7748 |
| 0.0955 | 4.9386 | 5.2183 | 0.2281 | 0.6113 | 0.6909 |
| 0.0981 | 4.8384 | 5.1265 | 0.2308 | 0.5529 | 0.6253 |
| 0.1008 | 4.7495 | 5.0517 | 0.2334 | 0.4963 | 0.5616 |
| 0.1034 | 4.6489 | 4.9568 | 0.2361 | 0.4416 | 0.4999 |
| 0.1061 | 4.5480 | 4.8605 | 0.2387 | 0.3770 | 0.4264 |
| 0.1088 | 4.4593 | 4.7821 | 0.2414 | 0.3270 | 0.3701 |
| 0.1114 | 4.3582 | 4.6834 | 0.2440 | 0.2793 | 0.3162 |
| 0.1141 | 4.2697 | 4.6030 | 0.2467 | 0.2340 | 0.2651 |
| 0.1167 | 4.1682 | 4.5019 | 0.2493 | 0.1914 | 0.2169 |
| 0.1194 | 4.0799 | 4.4197 | 0.2520 | 0.1426 | 0.1613 |
| 0.1220 | 3.9782 | 4.3165 | 0.2546 | 0.1068 | 0.1209 |
| 0.1247 | 3.8901 | 4.2327 | 0.2573 | 0.0748 | 0.0846 |
| 0.1273 | 3.7885 | 4.1279 | 0.2599 | 0.0467 | 0.0529 |
| 0.1300 | 3.7007 | 4.0428 | 0.2626 | 0.0235 | 0.0266 |

TABLE IV. Recoil spectrum from $\nu_e - e$ and $\nu_\mu - e$ scattering by $^8$B neutrinos. The kinetic energy of the electron is denoted by $T$ (measured in MeV) and the normalized probability distributions per MeV by $P(T)_{\nu_e}$ and $P(T)_{\nu_\mu}$.

| $T$ | $P(T)_{\nu_e}$ | $P(T)_{\nu_\mu}$ | $T$ | $P(T)_{\nu_e}$ | $P(T)_{\nu_\mu}$ |
|---|---|---|---|---|---|
| 0.0000 | 0.1671 | 0.2101 | 7.8584 | 0.0467 | 0.0401 |
| 0.1541 | 0.1658 | 0.2051 | 8.0125 | 0.0440 | 0.0377 |
| 0.3082 | 0.1646 | 0.2006 | 8.1666 | 0.0414 | 0.0355 |
| 0.4623 | 0.1635 | 0.1962 | 8.3207 | 0.0388 | 0.0332 |
| 0.6163 | 0.1624 | 0.1919 | 8.4747 | 0.0363 | 0.0309 |
| 0.7704 | 0.1613 | 0.1878 | 8.6288 | 0.0339 | 0.0289 |
| 0.9245 | 0.1601 | 0.1838 | 8.7829 | 0.0316 | 0.0268 |
| 1.0786 | 0.1589 | 0.1799 | 8.9370 | 0.0293 | 0.0248 |
| 1.2327 | 0.1576 | 0.1761 | 9.0911 | 0.0271 | 0.0230 |
| 1.3868 | 0.1563 | 0.1724 | 9.2452 | 0.0250 | 0.0212 |
| 1.5409 | 0.1549 | 0.1687 | 9.3993 | 0.0230 | 0.0194 |
| 1.6949 | 0.1534 | 0.1651 | 9.5533 | 0.0210 | 0.0178 |
| 1.8490 | 0.1519 | 0.1616 | 9.7074 | 0.0193 | 0.0163 |
| 2.0031 | 0.1502 | 0.1581 | 9.8615 | 0.0175 | 0.0148 |
| 2.1572 | 0.1485 | 0.1546 | 10.0156 | 0.0158 | 0.0133 |
| 2.3113 | 0.1467 | 0.1512 | 10.1697 | 0.0144 | 0.0121 |
| 2.4654 | 0.1449 | 0.1479 | 10.3238 | 0.0129 | 0.0108 |
| 2.6195 | 0.1429 | 0.1445 | 10.4779 | 0.0115 | 0.0096 |
| 2.7736 | 0.1408 | 0.1411 | 10.6319 | 0.0102 | 0.0085 |
| 2.9276 | 0.1387 | 0.1377 | 10.7860 | 0.0090 | 0.0076 |
| 3.0817 | 0.1365 | 0.1345 | 10.9401 | 0.0079 | 0.0066 |
| 3.2358 | 0.1342 | 0.1311 | 11.0942 | 0.0069 | 0.0058 |
| 3.3899 | 0.1318 | 0.1278 | 11.2483 | 0.0060 | 0.0050 |
| 3.5440 | 0.1293 | 0.1245 | 11.4024 | 0.0052 | 0.0043 |
| 3.6981 | 0.1269 | 0.1213 | 11.5565 | 0.0044 | 0.0037 |
| 3.8521 | 0.1242 | 0.1180 | 11.7105 | 0.0037 | 0.0031 |
| 4.0062 | 0.1215 | 0.1147 | 11.8646 | 0.0031 | 0.0026 |
| 4.1603 | 0.1188 | 0.1114 | 12.0187 | 0.0026 | 0.0021 |
| 4.3144 | 0.1161 | 0.1082 | 12.1728 | 0.0021 | 0.0017 |
| 4.4685 | 0.1132 | 0.1049 | 12.3269 | 0.0017 | 0.0014 |
| 4.6226 | 0.1102 | 0.1017 | 12.4810 | 0.0013 | 0.0011 |
| 4.7767 | 0.1074 | 0.0985 | 12.6351 | 0.0010 | 0.0009 |
| 4.9308 | 0.1044 | 0.0953 | 12.7891 | 0.0008 | 0.0007 |
| 5.0848 | 0.1013 | 0.0921 | 12.9432 | 0.0006 | 0.0005 |
| 5.2389 | 0.0982 | 0.0888 | 13.0973 | 0.0004 | 0.0004 |
| 5.3930 | 0.0952 | 0.0858 | 13.2514 | 0.0003 | 0.0003 |
| 5.5471 | 0.0921 | 0.0826 | 13.4055 | 0.0002 | 0.0002 |
| 5.7012 | 0.0889 | 0.0795 | 13.5596 | 0.0001 | 0.0001 |
| 5.8553 | 0.0857 | 0.0763 | 13.7137 | 0.0001 | 0.0001 |
| 6.0094 | 0.0827 | 0.0734 | 13.8678 | 0.0001 | 0.0001 |
| 6.1634 | 0.0795 | 0.0703 | 14.0218 | 0.0000 | 0.0000 |
| 6.3175 | 0.0764 | 0.0673 | 14.1759 | 0.0000 | 0.0000 |
| 6.4716 | 0.0734 | 0.0645 | 14.3300 | 0.0000 | 0.0000 |
| 6.6257 | 0.0702 | 0.0615 | 14.4841 | 0.0000 | 0.0000 |
| 6.7798 | 0.0671 | 0.0586 | 14.6382 | 0.0000 | 0.0000 |
| 6.9339 | 0.0640 | 0.0558 | 14.7923 | 0.0000 | 0.0000 |
| 7.0880 | 0.0611 | 0.0531 | 14.9464 | 0.0000 | 0.0000 |
| 7.2420 | 0.0581 | 0.0504 | 15.1004 | 0.0000 | 0.0000 |
| 7.3961 | 0.0552 | 0.0477 | 15.2545 | 0.0000 | 0.0000 |
| 7.5502 | 0.0524 | 0.0452 | 0.0000 | 0.0000 | 0.0000 |
| 7.7043 | 0.0495 | 0.0426 | | | |

TABLE V. Recoil spectrum from $\nu_e - e$ and $\nu_\mu - e$ scattering by $^{13}$N neutrinos. The kinetic energy of the electron is denoted by $T$ (measured in MeV) and the normalized probability distribution per MeV by $P(T)_{\nu_e}$ and $P(T)_{\nu_\mu}$.

| $T$ | $P(T)_{\nu_e}$ | $P(T)_{\nu_\mu}$ | $T$ | $P(T)_{\nu_e}$ | $P(T)_{\nu_\mu}$ |
|---|---|---|---|---|---|
| 0.0000 | 2.1691 | 2.1421 | 0.5011 | 0.9824 | 0.9679 |
| 0.0098 | 2.1472 | 2.1324 | 0.5109 | 0.9539 | 0.9383 |
| 0.0197 | 2.1265 | 2.1210 | 0.5208 | 0.9253 | 0.9087 |
| 0.0295 | 2.1060 | 2.1079 | 0.5306 | 0.9033 | 0.8866 |
| 0.0393 | 2.0858 | 2.0941 | 0.5404 | 0.8748 | 0.8573 |
| 0.0491 | 2.0657 | 2.0791 | 0.5502 | 0.8462 | 0.8279 |
| 0.0590 | 2.0461 | 2.0646 | 0.5601 | 0.8177 | 0.7987 |
| 0.0688 | 2.0259 | 2.0478 | 0.5699 | 0.7959 | 0.7771 |
| 0.0786 | 2.0056 | 2.0300 | 0.5797 | 0.7676 | 0.7483 |
| 0.0884 | 1.9859 | 2.0133 | 0.5895 | 0.7393 | 0.7195 |
| 0.0983 | 1.9662 | 1.9960 | 0.5994 | 0.7111 | 0.6910 |
| 0.1081 | 1.9453 | 1.9757 | 0.6092 | 0.6898 | 0.6699 |
| 0.1179 | 1.9253 | 1.9571 | 0.6190 | 0.6620 | 0.6419 |
| 0.1277 | 1.9051 | 1.9381 | 0.6288 | 0.6343 | 0.6141 |
| 0.1376 | 1.8849 | 1.9185 | 0.6387 | 0.6068 | 0.5865 |
| 0.1474 | 1.8644 | 1.8984 | 0.6485 | 0.5862 | 0.5663 |
| 0.1572 | 1.8417 | 1.8742 | 0.6583 | 0.5592 | 0.5394 |
| 0.1670 | 1.8207 | 1.8530 | 0.6682 | 0.5324 | 0.5127 |
| 0.1769 | 1.7995 | 1.8312 | 0.6780 | 0.5058 | 0.4864 |
| 0.1867 | 1.7779 | 1.8090 | 0.6878 | 0.4862 | 0.4673 |
| 0.1965 | 1.7562 | 1.7864 | 0.6976 | 0.4603 | 0.4417 |
| 0.2063 | 1.7342 | 1.7632 | 0.7075 | 0.4347 | 0.4166 |
| 0.2162 | 1.7118 | 1.7397 | 0.7173 | 0.4158 | 0.3982 |
| 0.2260 | 1.6892 | 1.7158 | 0.7271 | 0.3910 | 0.3740 |
| 0.2358 | 1.6664 | 1.6914 | 0.7369 | 0.3666 | 0.3502 |
| 0.2456 | 1.6432 | 1.6666 | 0.7468 | 0.3427 | 0.3268 |
| 0.2555 | 1.6233 | 1.6467 | 0.7566 | 0.3252 | 0.3100 |
| 0.2653 | 1.5997 | 1.6214 | 0.7664 | 0.3021 | 0.2876 |
| 0.2751 | 1.5758 | 1.5957 | 0.7762 | 0.2796 | 0.2658 |
| 0.2849 | 1.5516 | 1.5695 | 0.7861 | 0.2631 | 0.2500 |
| 0.2948 | 1.5271 | 1.5431 | 0.7959 | 0.2418 | 0.2294 |
| 0.3046 | 1.5022 | 1.5163 | 0.8057 | 0.2209 | 0.2093 |
| 0.3144 | 1.4771 | 1.4892 | 0.8155 | 0.2006 | 0.1898 |
| 0.3242 | 1.4517 | 1.4619 | 0.8254 | 0.1861 | 0.1760 |
| 0.3341 | 1.4307 | 1.4405 | 0.8352 | 0.1671 | 0.1578 |
| 0.3439 | 1.4050 | 1.4128 | 0.8450 | 0.1488 | 0.1403 |
| 0.3537 | 1.3789 | 1.3847 | 0.8548 | 0.1358 | 0.1280 |
| 0.3636 | 1.3526 | 1.3564 | 0.8647 | 0.1189 | 0.1119 |
| 0.3734 | 1.3260 | 1.3279 | 0.8745 | 0.1028 | 0.0967 |
| 0.3832 | 1.2992 | 1.2992 | 0.8843 | 0.0875 | 0.0822 |
| 0.3930 | 1.2775 | 1.2772 | 0.8941 | 0.0770 | 0.0723 |
| 0.4029 | 1.2504 | 1.2483 | 0.9040 | 0.0634 | 0.0595 |
| 0.4127 | 1.2230 | 1.2191 | 0.9138 | 0.0508 | 0.0476 |
| 0.4225 | 1.1955 | 1.1898 | 0.9236 | 0.0424 | 0.0397 |
| 0.4323 | 1.1677 | 1.1604 | 0.9334 | 0.0317 | 0.0296 |
| 0.4422 | 1.1457 | 1.1381 | 0.9433 | 0.0222 | 0.0208 |
| 0.4520 | 1.1178 | 1.1086 | 0.9531 | 0.0163 | 0.0152 |
| 0.4618 | 1.0896 | 1.0791 | 0.9629 | 0.0092 | 0.0086 |
| 0.4716 | 1.0614 | 1.0494 | 0.9727 | 0.0037 | 0.0034 |
| 0.4815 | 1.0330 | 1.0197 | 0.9826 | 0.0012 | 0.0011 |
| 0.4913 | 1.0109 | 0.9975 | | | |

TABLE VI. Recoil spectrum from $\nu_e - e$ and $\nu_\mu - e$ scattering by $^{15}$O neutrinos. The kinetic energy of the electron is denoted by $T$ (measured in MeV) and the normalized probability distribution per MeV by $P(T)_{\nu_e}$ and $P(T)_{\nu_\mu}$.

| $T$ | $P(T)_{\nu_e}$ | $P(T)_{\nu_\mu}$ | $T$ | $P(T)_{\nu_e}$ | $P(T)_{\nu_\mu}$ |
|---|---|---|---|---|---|
| 0.0000 | 1.4034 | 1.4756 | 0.7654 | 0.6470 | 0.6182 |
| 0.0150 | 1.3907 | 1.4639 | 0.7804 | 0.6278 | 0.5986 |
| 0.0300 | 1.3787 | 1.4517 | 0.7954 | 0.6086 | 0.5791 |
| 0.0450 | 1.3668 | 1.4387 | 0.8104 | 0.5893 | 0.5596 |
| 0.0600 | 1.3551 | 1.4255 | 0.8254 | 0.5745 | 0.5449 |
| 0.0750 | 1.3436 | 1.4126 | 0.8404 | 0.5554 | 0.5258 |
| 0.0900 | 1.3319 | 1.3986 | 0.8555 | 0.5363 | 0.5066 |
| 0.1051 | 1.3204 | 1.3851 | 0.8705 | 0.5171 | 0.4876 |
| 0.1201 | 1.3084 | 1.3703 | 0.8855 | 0.4981 | 0.4687 |
| 0.1351 | 1.2968 | 1.3562 | 0.9005 | 0.4836 | 0.4546 |
| 0.1501 | 1.2852 | 1.3420 | 0.9155 | 0.4648 | 0.4361 |
| 0.1651 | 1.2726 | 1.3260 | 0.9305 | 0.4461 | 0.4178 |
| 0.1801 | 1.2606 | 1.3112 | 0.9455 | 0.4275 | 0.3996 |
| 0.1951 | 1.2485 | 1.2962 | 0.9605 | 0.4090 | 0.3817 |
| 0.2101 | 1.2362 | 1.2810 | 0.9755 | 0.3951 | 0.3683 |
| 0.2251 | 1.2237 | 1.2656 | 0.9905 | 0.3770 | 0.3508 |
| 0.2401 | 1.2111 | 1.2499 | 1.0055 | 0.3591 | 0.3336 |
| 0.2551 | 1.1983 | 1.2340 | 1.0205 | 0.3414 | 0.3166 |
| 0.2701 | 1.1852 | 1.2180 | 1.0355 | 0.3239 | 0.2998 |
| 0.2852 | 1.1719 | 1.2017 | 1.0506 | 0.3109 | 0.2876 |
| 0.3002 | 1.1584 | 1.1851 | 1.0656 | 0.2939 | 0.2714 |
| 0.3152 | 1.1447 | 1.1684 | 1.0806 | 0.2772 | 0.2555 |
| 0.3302 | 1.1307 | 1.1515 | 1.0956 | 0.2607 | 0.2400 |
| 0.3452 | 1.1164 | 1.1343 | 1.1106 | 0.2485 | 0.2286 |
| 0.3602 | 1.1020 | 1.1170 | 1.1256 | 0.2328 | 0.2137 |
| 0.3752 | 1.0872 | 1.0995 | 1.1406 | 0.2173 | 0.1992 |
| 0.3902 | 1.0722 | 1.0817 | 1.1556 | 0.2021 | 0.1850 |
| 0.4052 | 1.0570 | 1.0638 | 1.1706 | 0.1874 | 0.1712 |
| 0.4202 | 1.0415 | 1.0457 | 1.1856 | 0.1766 | 0.1613 |
| 0.4352 | 1.0258 | 1.0274 | 1.2006 | 0.1625 | 0.1482 |
| 0.4502 | 1.0098 | 1.0090 | 1.2156 | 0.1489 | 0.1356 |
| 0.4652 | 0.9965 | 0.9942 | 1.2307 | 0.1358 | 0.1235 |
| 0.4803 | 0.9802 | 0.9755 | 1.2457 | 0.1262 | 0.1147 |
| 0.4953 | 0.9636 | 0.9567 | 1.2607 | 0.1139 | 0.1034 |
| 0.5103 | 0.9467 | 0.9377 | 1.2757 | 0.1021 | 0.0925 |
| 0.5253 | 0.9297 | 0.9186 | 1.2907 | 0.0907 | 0.0821 |
| 0.5403 | 0.9124 | 0.8994 | 1.3057 | 0.0799 | 0.0722 |
| 0.5553 | 0.8949 | 0.8801 | 1.3207 | 0.0723 | 0.0653 |
| 0.5703 | 0.8808 | 0.8649 | 1.3357 | 0.0624 | 0.0563 |
| 0.5853 | 0.8630 | 0.8455 | 1.3507 | 0.0532 | 0.0479 |
| 0.6003 | 0.8450 | 0.8260 | 1.3657 | 0.0445 | 0.0401 |
| 0.6153 | 0.8269 | 0.8064 | 1.3807 | 0.0385 | 0.0347 |
| 0.6303 | 0.8085 | 0.7867 | 1.3957 | 0.0310 | 0.0279 |
| 0.6453 | 0.7900 | 0.7669 | 1.4107 | 0.0241 | 0.0217 |
| 0.6604 | 0.7754 | 0.7517 | 1.4258 | 0.0180 | 0.0161 |
| 0.6754 | 0.7568 | 0.7320 | 1.4408 | 0.0140 | 0.0125 |
| 0.6904 | 0.7380 | 0.7122 | 1.4558 | 0.0091 | 0.0082 |
| 0.7054 | 0.7191 | 0.6924 | 1.4708 | 0.0051 | 0.0046 |
| 0.7204 | 0.7000 | 0.6727 | 1.4858 | 0.0020 | 0.0018 |
| 0.7354 | 0.6810 | 0.6529 | 1.5008 | 0.0007 | 0.0006 |
| 0.7504 | 0.6661 | 0.6379 | | | |

TABLE VII. Scattering cross sections for individual neutrino energies. The tabulated values were determined for $T_{min} = 0.0$ MeV. The neutrino energy, $q$, is given in MeV and the cross sections, $\sigma_{\nu_e} - e$ and $\sigma_{\nu_\mu} - e$, are given in units of $10^{-46}$ cm$^2$.

| $q$ | $\sigma_{\nu_e} - e$ | $\sigma_{\nu_\mu} - e$ | $q$ | $\sigma_{\nu_e} - e$ | $\sigma_{\nu_\mu} - e$ |
|---|---|---|---|---|---|
| 0.38 | 1.92e+01 | 5.09e+00 | 12.00 | 1.08e+03 | 1.90e+02 |
| 0.86 | 5.79e+01 | 1.28e+01 | 14.00 | 1.27e+03 | 2.21e+02 |
| 1.00 | 6.98e+01 | 1.50e+01 | 16.00 | 1.45e+03 | 2.53e+02 |
| 1.44 | 1.09e+02 | 2.21e+01 | 18.00 | 1.64e+03 | 2.85e+02 |
| 2.00 | 1.59e+02 | 3.11e+01 | 20.00 | 1.82e+03 | 3.16e+02 |
| 3.00 | 2.51e+02 | 4.70e+01 | 25.00 | 2.28e+03 | 3.95e+02 |
| 4.00 | 3.42e+02 | 6.29e+01 | 30.00 | 2.74e+03 | 4.74e+02 |
| 5.00 | 4.35e+02 | 7.87e+01 | 40.00 | 3.67e+03 | 6.32e+02 |
| 7.00 | 6.19e+02 | 1.10e+02 | 50.00 | 4.59e+03 | 7.90e+02 |
| 10.00 | 8.96e+02 | 1.58e+02 | 60.00 | 5.52e+03 | 9.48e+02 |

TABLE VIII. Total neutrino-electron scattering cross sections. Radiative corrections were included and $\sin^2 \hat{\theta}_W = 0.2317$ was used. The minimum allowed recoil kinetic energy is zero in all cases considered in this table; the maximum recoil energy is given in column 3. The neutrino energy, $q$, and the maximum electron recoil energy, $T_{\max}$, are given in MeV; the neutrino cross sections, $\sigma_{\nu_e-e}$ and $\sigma_{\nu_\mu-e}$ are given in units of $10^{-46}$ cm$^2$.

| Source | $q$ | $T_{\max}$ | $\sigma_{\nu_e} - e$ | $\sigma_{\nu_\mu} - e$ |
|---|---|---|---|---|
| $pp$ | $\leq 0.420$ | 0.261 | 11.6 | 3.3 |
| $^7$Be | 0.862 | 0.665 | 57.9 | 12.8 |
| $^7$Be | 0.384 | 0.231 | 19.2 | 5.1 |
| $^8$B | $< 15.0$ | 14.5 | 594 | 106 |
| $^{13}$N | $\leq 1.199$ | 0.988 | 45.8 | 10.4 |
| $^{15}$O | $\leq 1.1732$ | 1.509 | 70.8 | 15.1 |